\documentclass[reprint,aps,pra,longbibliography,floatfix]{revtex4-2}
\usepackage{graphicx}
\usepackage{dcolumn}% Align table columns on decimal point
\usepackage{bm}% bold math
\usepackage[dvipsnames]{xcolor}
\usepackage{color}
\usepackage[colorlinks,citecolor=blue,linkcolor=blue,urlcolor=blue,linktocpage]{hyperref}
\usepackage{textcomp}
\definecolor{refcolor}{RGB}{0,0,255}
\usepackage{amsmath,amssymb,amsthm,amsfonts}
\usepackage{comment}
\usepackage{tikz}
\usetikzlibrary{shapes.geometric, arrows.meta, positioning, calc}

\begin{document}

\title{Shapiro steps of superfluid Fermi gases in a ring trap across the BCS-BEC crossover}% Force line breaks with \\

\author{Hikaru Kuriki}
\email{1225523@ed.tus.ac.jp}
\author{Masaya Kunimi}%
\email{kunimi@rs.tus.ac.jp}
\author{Tetsuro Nikuni}%
\affiliation{Department of Physics, Tokyo University of Science, 1-3 Kagurazaka, Tokyo 162-8601, Japan}%Lines break automatically or can be forced with \\

\date{\today}% It is always \today, today,

\begin{abstract} \label{abstract}
  We investigate the transport properties of a superfluid Fermi gas confined in a ring trap with a moving potential barrier across the Bardeen-Cooper-Schrieffer to Bose-Einstein condensate crossover.
  Employing time-dependent Bogoliubov--de Gennes (BdG) equations, we simulate the dynamics of a Josephson junction biased by both dc and ac currents.
  Over a wide range of interaction strengths, the barrier-velocity dependence of the chemical-potential difference exhibits low-order plateau structures, consistent with Shapiro steps,  with fitted levels
close to integer multiples of $\hbar\omega/2$ within the phase-coherent regime.
  This factor of $1/2$ reflects our convention of defining the chemical potential per single fermion in the BdG framework.
  Microscopic analysis reveals that these fundamental steps originate from synchronized phase slips mediated by periodic soliton generation at the barrier. 
 Our findings clarify the role of interaction regimes in the nonequilibrium phase dynamics of ring-trapped fermionic superfluids and provide microscopic insights relevant to future studies of atomtronic systems with nontrivial topology.
\end{abstract}

\maketitle

%%%%%%%%%%%%%%%%%%%%%%%%%%%%%%%%%%%%%%%%%%%%%%%%%%%%%%%%%%%%%%%%%%%%%%%%%%%%%%%%%%%%%%%%%%%%%%%%%%%%%%
%%%%%%%%%%%%%%%%%%%%%%%%%%%%%%%%%%%%%%%%%%%%%%%%%%%%%%%%%%%%%%%%%%%%%%%%%%%%%%%%%%%%%%%%%%%%%%%%%%%%%
%%%%%%%%%%%%%%%%%%%%%%%%%%%%%%%%%%%%%%%%%%%%%%%%%%%%%%%%%%%%%%%%%%%%%%%%%%%%%%%%%%%%%%%%%%%%%%%%%%%%%%
\section{Introduction} \label{sec:intro}
%%%%%%%%%%%%%%%%%%%%%%%%%%%%%%%%%%%%%%%%%%%%%%%%%%%%%%%%%%%%%%%%%%%%%%%%%%%%%%%%%%%%%%%%%%%%%%%%%%%%%%
%%%%%%%%%%%%%%%%%%%%%%%%%%%%%%%%%%%%%%%%%%%%%%%%%%%%%%%%%%%%%%%%%%%%%%%%%%%%%%%%%%%%%%%%%%%%%%%%%%%%%%
%%%%%%%%%%%%%%%%%%%%%%%%%%%%%%%%%%%%%%%%%%%%%%%%%%%%%%%%%%%%%%%%%%%%%%%%%%%%%%%%%%%%%%%%%%%%%%%%%%%%%%

Ultracold atomic gases confined in ring traps have played a pivotal role in the study of superfluidity~\cite{ryuObservationPersistentFlow2007a,ramanathanSuperflowToroidalBoseEinstein2011,moulderQuantizedSupercurrentDecay2012,beattiePersistentCurrentsSpinor2013,neelyCharacteristicsTwoDimensionalQuantum2013,wrightDrivingPhaseSlips2013,wrightThresholdCreatingExcitations2013,ryuExperimentalRealizationJosephson2013,ryuCreationMatterWave2014,eckelHysteresisQuantizedSuperfluid2014,jendrzejewskiResistiveFlowWeakly2014,eckelInterferometricMeasurementCurrentPhase2014,cormanQuenchInducedSupercurrentsAnnular2014,kumarTemperatureinducedDecayPersistent2017,pandeyHypersonicBoseEinstein2019,kunimiDecayMechanismsSuperflow2019,ryuQuantumInterferenceCurrents2020,pecciProbingBCSBECCrossover2021,ogrenStationaryStatesBose2021,caiPersistentCurrentsRings2022,delpaceImprintingPersistentCurrents2022,xhaniDecayPersistentCurrents2023,pezzeStabilizingPersistentCurrents2024c,fernandezAngularMomentumRotating2025a,ganJosephsonDynamics2D2025,pradhanFractionalShapiroSteps2025a,chenDynamicGenerationSuperflow2025b,xhaniStabilityPersistentCurrents2025,xhaniTuningCriticalCurrent2025,tuzemenPairbreakingFundamentalLimit2025,ciszakCooperativeStabilizationPersistent2026a,ragoleInteractingAtomicInterferometry2016,poloQuantumSolitonsAtomtronic2021}. Their pristine nature and high degree of controllability have enabled the exploration of fundamental superfluid properties, ranging from persistent currents~\cite{ramanathanSuperflowToroidalBoseEinstein2011,moulderQuantizedSupercurrentDecay2012,beattiePersistentCurrentsSpinor2013,eckelInterferometricMeasurementCurrentPhase2014,cormanQuenchInducedSupercurrentsAnnular2014,kumarTemperatureinducedDecayPersistent2017,caiPersistentCurrentsRings2022,delpaceImprintingPersistentCurrents2022,pezzeStabilizingPersistentCurrents2024c} and critical velocities~\cite{wrightDrivingPhaseSlips2013,wrightThresholdCreatingExcitations2013} to hysteresis loops~\cite{eckelHysteresisQuantizedSuperfluid2014}, as well as Josephson junction physics~\cite{ganJosephsonDynamics2D2025}. Furthermore, drawing an analogy to superconducting quantum interference devices~\cite{clarkeSQUIDHandbookFundamentals2004}, these systems have attracted growing attention as atomtronic quantum interference devices~\cite{ryuExperimentalRealizationJosephson2013,ragoleInteractingAtomicInterferometry2016,poloQuantumSolitonsAtomtronic2021,amicoRoadmapAtomtronicsState2021,amicoColloquiumAtomtronicCircuits2022,gorgRealizingAtomicQuantum2025}, which are expected to serve as fundamental building blocks for atomtronic circuits.

In particular, recent experimental realizations of ring traps for fermionic superfluids~\cite{caiPersistentCurrentsRings2022,delpaceImprintingPersistentCurrents2022} have stimulated intense theoretical~\cite{pecciProbingBCSBECCrossover2021,xhaniDecayPersistentCurrents2023,chenDynamicGenerationSuperflow2025b,xhaniStabilityPersistentCurrents2025,tuzemenPairbreakingFundamentalLimit2025} and experimental~\cite{fernandezAngularMomentumRotating2025a} investigations. Ultracold Fermi gases offer a unique advantage due to their highly tunable interactions, which allow the exploration of diverse superfluid properties covering the full range from the weak-coupling Bardeen-Cooper-Schrieffer (BCS) regime to the strong-coupling Bose-Einstein condensate (BEC) regime, including the intermediate unitary Fermi gas (UFG) regime where the $s$-wave scattering length diverges. Consequently, these systems are expected to exhibit richer and more complex physics than their bosonic counterparts, such as pair-breaking excitations that become particularly important in the BCS regime~\cite{xhaniStabilityPersistentCurrents2025,tuzemenPairbreakingFundamentalLimit2025} and tunneling transport behavior in the unitary regime~\cite{delpaceTunnelingTransportUnitary2021}. 

One of the most fundamental phenomena in superfluidity is the Josephson effect~\cite{josephsonPossibleNewEffects1962b}, which occurs in Josephson junctions (JJs)---systems consisting of two reservoirs weakly coupled by a potential barrier. In such systems, analogous to superconducting JJs, the transition from the dc to the ac Josephson effect~\cite{albiezDirectObservationTunneling2005,levyAcDcJosephson2007,kwonStronglyCorrelatedSuperfluid2020} and current-phase relations~\cite{eckelInterferometricMeasurementCurrentPhase2014,valtolinaJosephsonEffectFermionic2015a, burchiantiConnectingDissipationPhase2018,luickIdealJosephsonJunction2020,delpaceTunnelingTransportUnitary2021} have been experimentally confirmed in ultracold atomic systems. One of the methods used to realize the Josephson effect is the motion of a potential barrier separating two ultracold atomic gases at constant velocity~\cite{giovanazziJosephsonEffectsDilute2000,levyAcDcJosephson2007,kwonStronglyCorrelatedSuperfluid2020,delpaceTunnelingTransportUnitary2021}. In this setup, the analogy to the Josephson effect is established as follows: when the barrier velocity is below a critical value, the tunneling current remains coherent, and no chemical-potential difference (analogous to voltage) arises between the two regions. In contrast, when the barrier velocity exceeds the critical value, a finite chemical-potential difference emerges. 

This transport behavior in JJs becomes particularly rich when a periodic modulation is superimposed on the constant barrier motion~\cite{singhShapiroStepsDriven2024b,pradhanFractionalShapiroSteps2025a,singhControlledGeneration3D2025a}. In superconducting JJs driven by microwave radiation, such modulation facilitates photon-assisted tunneling processes, causing the phase of Cooper pairs to synchronize with the external ac drive~\cite{baronePhysicsApplicationsJosephson1982}. This synchronization manifests as discrete plateaus in the current-voltage characteristics, widely referred to as Shapiro steps~\cite{shapiroJosephsonCurrentsSuperconducting1963,grimesMillimeterWaveMixingJosephson1968}. Recently, an analogous phenomenon has been demonstrated in ultracold atomic systems. In experiments utilizing box traps, distinct Shapiro steps at integer multiples of $\hbar\omega$ were observed in the response of the chemical-potential difference against the dc barrier velocity for both bosonic~\cite{bernhartObservationShapiroSteps2025a} and fermionic~\cite{delpaceShapiroStepsStronglyinteracting2025a} superfluids, where $\omega$ denotes the frequency of the modulation drive.

Despite these successful observations in box potentials, the realization and microscopic understanding of Shapiro steps in ring-trapped fermionic superfluids across the BCS-BEC crossover remain largely unexplored. 
In ring systems, the phase winding number is well-defined, and several metastable states characterized by the winding number exist. Furthermore, a comprehensive numerical investigation of these dynamics remains absent. This is especially critical for Fermi gases, where the behavior of Shapiro steps across the entire BCS-BEC crossover is nontrivial. The complex interplay between the ac driving and the internal degrees of freedom---such as pair-breaking excitations in the BCS regime---cannot be fully captured by simple phenomenological models. Therefore, a microscopic theoretical study is essential to elucidate the nonequilibrium dynamics of Shapiro steps in fermionic superfluids in this topologically nontrivial geometry.

In this paper, we present a microscopic study of Shapiro steps in a superfluid Fermi gas confined in a ring trap with a moving potential barrier. We model the barrier by a Gaussian potential whose motion contains both a constant (dc) component and a periodic (ac) modulation, which effectively realizes a Josephson junction biased by combined dc and ac currents. Starting from a self-consistent stationary state obtained from the Bogoliubov--de Gennes (BdG) equations, the gap equation, and the particle number equation, we compute the subsequent dynamics by solving the time-dependent BdG (tdBdG) equations. We show that the effective dc chemical-potential difference exhibits low-order plateau structures whose fitted levels lie
close to integer multiples of $\hbar\omega/2$, and we identify their microscopic origin in periodic phase slips mediated by soliton generation near the barrier.

This paper is organized as follows: In Sec.~\ref{sec:model}, we introduce our model of a ring-trapped Fermi superfluid with a moving Gaussian barrier and describe the numerical methods based on BdG and tdBdG equations. In Sec.~\ref{sec:results}, we present the current--chemical-potential characteristics and demonstrate Shapiro steps under dc and ac driving. We also provide a microscopic analysis of the order-parameter dynamics. Finally, Sec.~\ref{sec:conclusion} summarizes our findings and discusses future directions.

%%%%%%%%%%%%%%%%%%%%%%%%%%%%%%%%%%%%%%%%%%%%%%%%%%%%%%%%%%%%%%%%%%%%%%%%%%%%%%%%%%%%%%%%%%%%%%%%%%%%%%
%%%%%%%%%%%%%%%%%%%%%%%%%%%%%%%%%%%%%%%%%%%%%%%%%%%%%%%%%%%%%%%%%%%%%%%%%%%%%%%%%%%%%%%%%%%%%%%%%%%%%%
%%%%%%%%%%%%%%%%%%%%%%%%%%%%%%%%%%%%%%%%%%%%%%%%%%%%%%%%%%%%%%%%%%%%%%%%%%%%%%%%%%%%%%%%%%%%%%%%%%%%%%
\section{Model and Methods} \label{sec:model}
%%%%%%%%%%%%%%%%%%%%%%%%%%%%%%%%%%%%%%%%%%%%%%%%%%%%%%%%%%%%%%%%%%%%%%%%%%%%%%%%%%%%%%%%%%%%%%%%%%%%%%
%%%%%%%%%%%%%%%%%%%%%%%%%%%%%%%%%%%%%%%%%%%%%%%%%%%%%%%%%%%%%%%%%%%%%%%%%%%%%%%%%%%%%%%%%%%%%%%%%%%%%%
%%%%%%%%%%%%%%%%%%%%%%%%%%%%%%%%%%%%%%%%%%%%%%%%%%%%%%%%%%%%%%%%%%%%%%%%%%%%%%%%%%%%%%%%%%%%%%%%%%%%%%

We consider a superfluid Fermi gas confined in a ring trap at zero temperature with equal populations of the two spin components. The system is three-dimensional, and the position is denoted by $\bm{r} = (x,y,z)$. For simplicity, we allow spatial inhomogeneity only in the ring direction $x$ and impose periodic boundary conditions along all directions. This setup should be regarded as an effective one-dimensional ring geometry with transverse momentum modes.

%%%%%%%%%%%%%%%%%%%%%%%%%%%%%%%%%%%%%%%%%%%%%%%%%%%%%%%%%%%%%%%%%%%%%%%%%%%%%%%%%%%%%%%%%%%%%%%%%%%%%%%%%
%%%%%%%%%%%%%%%%%%%%%%%%%%%%%%%%%%%%%%%%%%%%%%%%%%%%%%%%%%%%%%%%%%%%%%%%%%%%%%%%%%%%%%%%%%%%%%%%%%%%%%%%%
%%%%%%%%%%%%%%%%%%%%%%%%%%%%%%%%%%%%%%%%%%%%%%%%%%%%%%%%%%%%%%%%%%%%%%%%%%%%%%%%%%%%%%%%%%%%%%%%%%%%%%%%%
\subsection{Potential barrier and chemical-potential difference} \label{sec:potential_barrier}
%%%%%%%%%%%%%%%%%%%%%%%%%%%%%%%%%%%%%%%%%%%%%%%%%%%%%%%%%%%%%%%%%%%%%%%%%%%%%%%%%%%%%%%%%%%%%%%%%%%%%%%%%
%%%%%%%%%%%%%%%%%%%%%%%%%%%%%%%%%%%%%%%%%%%%%%%%%%%%%%%%%%%%%%%%%%%%%%%%%%%%%%%%%%%%%%%%%%%%%%%%%%%%%%%%%
%%%%%%%%%%%%%%%%%%%%%%%%%%%%%%%%%%%%%%%%%%%%%%%%%%%%%%%%%%%%%%%%%%%%%%%%%%%%%%%%%%%%%%%%%%%%%%%%%%%%%%%%%

\begin{figure}[htbp]
  \centering
  \includegraphics[width=0.5\linewidth]{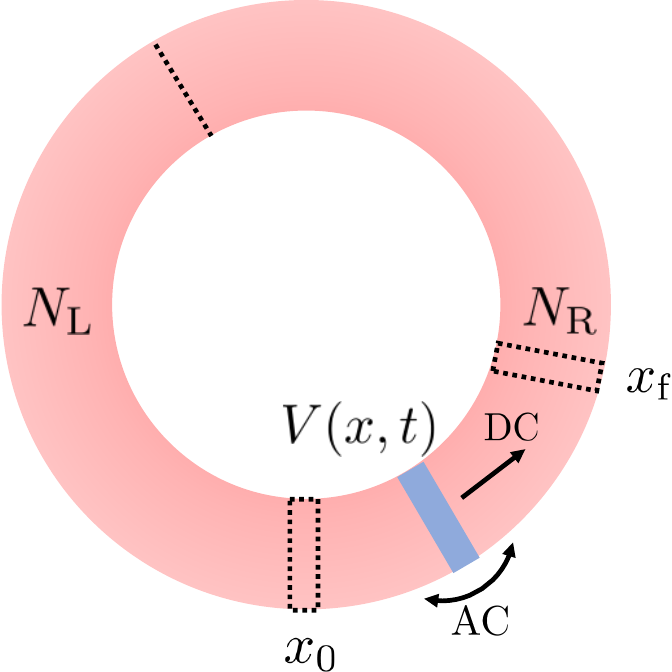}
  \caption{Schematic illustration of a ring-trapped Fermi superfluid with a moving Gaussian potential barrier $V(x,t)$ under combined dc and ac driving (blue shaded area). The barrier position $x_{\mathrm{b}}(t)$ is defined in Eq.~\eqref{eq:barrier_position}. $x_0$ is the initial position of the barrier, and $x_{\mathrm{f}} = x_{\mathrm{b}}(t_{\mathrm{f}})$ is the final position at the end of the simulation. The point opposite to the barrier position is used as an effective boundary to define the left and right reservoirs, which are separated by the barrier and its antipode (black dotted line). $N_{\mathrm{L}}$ and $N_{\mathrm{R}}$ represent the atom numbers in the left and right reservoirs, respectively.} 
  \label{fig:schematic_ring}
\end{figure}

We consider the dynamics of the system with a Gaussian potential barrier moving along the ring direction $x$ as
\begin{align}
  V(x,t) = V_0 \exp\left\{-2\frac{[x - x_{\mathrm{b}}(t)]^2}{w^2}\right\}, \label{eq:barrier_potential}
\end{align}
where $V_0$ and $w$ are the height and width of the barrier, respectively. The barrier position $x_{\mathrm{b}}(t)$ is given by
\begin{align}
  x_{\mathrm{b}}(t) = x_0 + v_{\mathrm{DC}} t + x_{\mathrm{AC}} \sin(\omega t), \label{eq:barrier_position}
\end{align}
where $x_0$ is the initial position of the barrier, $v_{\mathrm{DC}}$ is the dc velocity component, and $x_{\mathrm{AC}}$ and $\omega$ are the amplitude and frequency of the ac modulation, respectively. Here, the distance $|x-x_{\mathrm{b}}|$ is defined as the shortest periodic distance on the ring. This barrier motion effectively realizes a JJ biased by combined dc and ac currents.

Given the ring geometry, a single potential barrier does not divide the system into two separate reservoirs. To define ``left'' and ``right'' regions, we therefore introduce the point opposite to the barrier as an effective boundary and divide the ring into two domains separated by the barrier and its antipode (see Fig.~\ref{fig:schematic_ring} for a schematic illustration). We refer to this point as the edge, defined as $x_{\mathrm{edge}}(t) = \mathrm{mod}\left[ L_x/2 + x_{\mathrm{b}}(t), L_x\right]$, where $L_x$ is the system size along the ring direction $x$.

To quantify the transport properties, we first calculate the particle numbers in the left and right reservoirs, 
$N_{\mathrm{L}}(t)$ and $N_{\mathrm{R}}(t)$, respectively. 
We then define the population imbalance as
\begin{align}
z(t)=\frac{N_{\mathrm{R}}(t)-N_{\mathrm{L}}(t)}{N},\label{eq:definition_of_population_imbalance}
\end{align}
where $N=N_{\mathrm{L}}(t)+N_{\mathrm{R}}(t)$ denotes the total particle number. Then, we compute the chemical-potential difference between the two reservoirs as
\begin{align}
  \Delta \mu(t) = \frac{NE_{\mathrm{C}}}{2} z(t), \label{eq:chemical_potential_difference}
\end{align}
where $E_{\mathrm{C}}$ is the effective charging energy defined by
$E_{\mathrm{C}}=4\partial\mu/\partial N$. This quantity is related to the inverse compressibility of the reservoirs, which characterizes their density response to currents~\cite{giovanazziJosephsonEffectsDilute2000,meierJosephsonTunnelingWeakly2001,kwonStronglyCorrelatedSuperfluid2020,delpaceTunnelingTransportUnitary2021,singhShapiroStepsDriven2024b}. Here, the chemical potential $\mu$ is calculated as described in Sec.~\ref{sec:bdg_equations}. We evaluate $E_{\mathrm{C}}$ by numerically differentiating $\mu$ with respect to $N$ around the total particle number $N$ by solving the BdG equations introduced in Sec.~\ref{sec:bdg_equations} for various particle numbers (see Appendix~\ref{sec:charging_energy} for details).

We calculate $\Delta \mu (t)$ for various values of $v_{\mathrm{DC}}$. To extract the effective dc component from the transient ac oscillations, we follow the protocol utilized in recent theoretical~\cite{singhShapiroStepsDriven2024b} and experimental~\cite{delpaceShapiroStepsStronglyinteracting2025a} studies. Specifically, we fit the time evolution of $\Delta \mu(t)$ with a linear function, $\Delta \mu_{\mathrm{fit}}(t) = a_{\mathrm{fit}} t$, and evaluate this fitted function at the final simulation time $t_{\mathrm{f}}$. We identify this terminal value, $\Delta \mu_{\mathrm{fit}}(t_{\mathrm{f}})$, as the  finite-time effective dc chemical-potential difference, and denote it simply by $\Delta \mu (t_{\mathrm{f}})$ in the following. This evaluation method emulates the experimental measurement procedure, where the global chemical potential bias is extracted from a single \textit{in situ} density snapshot taken immediately after a finite period of barrier motion. This approach is physically justified provided that $t_{\mathrm{f}}$ is carefully chosen within an optimal time window: it must be long enough for the local density modulations to propagate across the system, yet short enough to prevent these emitted sound waves from circumnavigating the closed ring geometry and interfering with the barrier. Within this transient regime, the global density imbalance at $t_{\mathrm{f}}$ reflects the cumulative history of the local density modulations. Consequently, the time-integrated local dynamics are spatially mapped onto the global macroscopic observable at $t_{\mathrm{f}}$, allowing the terminal value of the fitted linear trend to provide a finite-time estimate of the effective dc component (see the Supplemental Material of Ref.~\cite{delpaceShapiroStepsStronglyinteracting2025a} for details). Provided that an appropriate finite observation time is chosen (e.g., a few driving cycles, as used below), the emitted density waves do not yet substantially disturb the density distribution over the entire ring, even if they have propagated to the antipodal region. Under such conditions, defining the left and right reservoirs using the antipodal point is expected to provide a reasonable measure of the population imbalance. Within this finite-time window, the extracted population imbalance is therefore not expected to be sensitive to small displacements of the antipodal boundary.

%%%%%%%%%%%%%%%%%%%%%%%%%%%%%%%%%%%%%%%%%%%%%%%%%%%%%%%%%%%%%%%%%%%%%%%%%%%%%%%%%%%%%%%%%%%%%%%%%%%%%%%%%
%%%%%%%%%%%%%%%%%%%%%%%%%%%%%%%%%%%%%%%%%%%%%%%%%%%%%%%%%%%%%%%%%%%%%%%%%%%%%%%%%%%%%%%%%%%%%%%%%%%%%%%%%
%%%%%%%%%%%%%%%%%%%%%%%%%%%%%%%%%%%%%%%%%%%%%%%%%%%%%%%%%%%%%%%%%%%%%%%%%%%%%%%%%%%%%%%%%%%%%%%%%%%%%%%%%
\subsection{BdG equations for the initial state} \label{sec:bdg_equations}
%%%%%%%%%%%%%%%%%%%%%%%%%%%%%%%%%%%%%%%%%%%%%%%%%%%%%%%%%%%%%%%%%%%%%%%%%%%%%%%%%%%%%%%%%%%%%%%%%%%%%%%%%
%%%%%%%%%%%%%%%%%%%%%%%%%%%%%%%%%%%%%%%%%%%%%%%%%%%%%%%%%%%%%%%%%%%%%%%%%%%%%%%%%%%%%%%%%%%%%%%%%%%%%%%%%
%%%%%%%%%%%%%%%%%%%%%%%%%%%%%%%%%%%%%%%%%%%%%%%%%%%%%%%%%%%%%%%%%%%%%%%%%%%%%%%%%%%%%%%%%%%%%%%%%%%%%%%%%

As the initial state, we first obtain a stationary solution of the BdG equations~\cite{giorginiTheoryUltracoldAtomic2008,antezzaDarkSolitonsSuperfluid2007a,degennesSuperconductivityMetalsAlloys2018} at $t=0$
\begin{align}
  \begin{pmatrix}
    h(\bm{r}) & \Delta(\bm{r}) \\
    \Delta^*(\bm{r}) & -h^*(\bm{r})
  \end{pmatrix}
  \begin{pmatrix}
    u_{\nu}(\bm{r}) \\
    v_{\nu}(\bm{r})
  \end{pmatrix}
  = E_{\nu}
  \begin{pmatrix}
    u_{\nu}(\bm{r}) \\
    v_{\nu}(\bm{r})
  \end{pmatrix},
  \label{eq:BdG}
\end{align}
where $\nu$ denotes the label of the eigenstate, the functions $u_\nu(\bm{r})$ and $v_\nu(\bm{r})$ are the quasiparticle wave functions and $E_\nu$ is the corresponding eigenenergy,  $h(\bm{r})=-\hbar^2 \nabla^2/(2m) + V(x, t=0) - \mu$ is a single-particle Hamiltonian for atoms of mass $m$ subject to the potential barrier $V(x, t=0)$ and $\mu$ is the chemical potential. The quasiparticle wave functions satisfy the normalization condition $\int d\bm{r} [u_\nu ^*(\bm{r})u_
{\nu'}(\bm{r}) + v_{\nu}^*(\bm{r}) v_{\nu'}(\bm{r})] = \delta_{\nu,\nu'}$. The order parameter $\Delta(\bm{r})$ is calculated self-consistently from the gap equation as
\begin{align}
  \Delta(\bm{r}) = -g \sum_{0\le E_\nu<E_{\mathrm{cut}}} u_{\nu}(\bm{r}) v_{\nu}^*(\bm{r}), \label{eq:gap_eq_static}
\end{align}
where $g$ is the coupling constant given by~\cite{giorginiTheoryUltracoldAtomic2008} 
\begin{align}
  \frac{1}{k_{\mathrm{F}}a} = \frac{8\pi E_{\mathrm{F}}}{gk_{\mathrm{F}}^{3}} + \frac{2}{\pi}\sqrt{\frac{E_{\mathrm{cut}}}{E_{\mathrm{F}}}}. \label{eq:coupling_const}
\end{align}
Here, $a$ is the $s$-wave scattering length characterizing the interaction between atoms of different spins, while $E_{\mathrm{F}} = \hbar^2 k_{\mathrm{F}}^2/(2m)$ and $k_{\mathrm{F}} = (3\pi^2 n_0)^{1/3}$ are the Fermi energy and Fermi wave number of an ideal Fermi gas of average density $n_0 = N/\mathcal{V}$ with system volume $\mathcal{V}$. We note that $1/(k_{\rm F}a)<0$, $1/(k_{\rm F}a)=0$, and $1/(k_{\rm F}a)>0$ correspond to the BCS, UFG, and BEC regimes, respectively. The summation in Eq.~\eqref{eq:gap_eq_static} is taken over all non-negative eigenenergies below a cutoff energy $E_{\mathrm{cut}}$ to remove the ultraviolet divergence in the BdG equations with the contact potential.  
We also calculate the particle number equation
\begin{align}
  N = 2 \sum_{E_\nu \ge 0} \int d\bm{r}~ |v_{\nu}(\bm{r})|^2, \label{eq:particle_number}
\end{align}
to determine the chemical potential $\mu$ self-consistently for a given total particle number $N$.

We note that the BdG approach employed here is a mean-field theory. While it provides a unified microscopic description across the BCS-BEC crossover, quantitative predictions near unitarity and toward the BEC regime should be interpreted with caution because strong-correlation and pairing-fluctuation effects beyond mean field are not fully captured~\cite{giorginiTheoryUltracoldAtomic2008}. Accordingly, our comparison across the crossover is intended primarily to characterize qualitative trends in the nonequilibrium dynamics.

Assuming spatial inhomogeneity solely along the ring direction $x$, we write the quasiparticle wave functions as $u_{\nu}(\bm{r}) = u_{\nu}(x) e^{i(k_y y+ k_z z)}/L_{\perp}$ and $v_{\nu}(\bm{r}) = v_\nu(x) e^{i(k_y y+ k_z z)}/L_{\perp}$, in which $k_y$ and $k_z$ are quantized according to $k_y = 2\pi \alpha_y/L_{\perp}$ and $k_z = 2\pi \alpha_z/L_{\perp}$ with integers $\alpha_y$ and $\alpha_z$, and $L_{\perp}$ is the system size in the transverse directions.
Then, Eqs.~\eqref{eq:BdG}, \eqref{eq:gap_eq_static}, and \eqref{eq:particle_number} reduce to effective one-dimensional forms along the $x$ direction with summations over the transverse modes $(k_y,k_z)$ as
\begin{gather}
  \begin{pmatrix}
    h(x) & \Delta(x) \\
    \Delta^*(x) & -h^{\ast}(x)
  \end{pmatrix}
  \begin{pmatrix}
    u_{\nu}(x) \\
    v_{\nu}(x)
  \end{pmatrix}
  = E_{\nu}
  \begin{pmatrix}
    u_{\nu}(x) \\
    v_{\nu}(x)
  \end{pmatrix}, \label{eq:BdG_1D} \\
  \Delta(x) = -\frac{g}{L_{\perp}^2} \sum_{k_y,k_z} \sum_{0\le E_\nu<E_{\mathrm{cut}}} u_{\nu}(x) v_{\nu}^*(x), \label{eq:gap_eq_static_1D} \\
  N = 2 \sum_{k_y,k_z} \sum_{E_{\nu} \ge 0} \int dx~ |v_{\nu}(x)|^2, \label{eq:particle_number_1D}
\end{gather}
where $h(x) = -\hbar^2(\partial^2/\partial x^2 - k_y^2 - k_z^2)/(2m) + V(x) - \mu$.

We numerically solve Eqs.~\eqref{eq:BdG_1D}, (\ref{eq:gap_eq_static_1D}), and \eqref{eq:particle_number_1D} self-consistently using the modified Broyden's method~\cite{johnsonModifiedBroydensMethod1988} as explained in Appendix~\ref{sec:modified_broyden}.

%%%%%%%%%%%%%%%%%%%%%%%%%%%%%%%%%%%%%%%%%%%%%%%%%%%%%%%%%%%%%%%%%%%%%%%%%%%%%%%%%%%%%%%%%%%%%%%%%%%%%%%%%
%%%%%%%%%%%%%%%%%%%%%%%%%%%%%%%%%%%%%%%%%%%%%%%%%%%%%%%%%%%%%%%%%%%%%%%%%%%%%%%%%%%%%%%%%%%%%%%%%%%%%%%%%
%%%%%%%%%%%%%%%%%%%%%%%%%%%%%%%%%%%%%%%%%%%%%%%%%%%%%%%%%%%%%%%%%%%%%%%%%%%%%%%%%%%%%%%%%%%%%%%%%%%%%%%%%
\subsection{tdBdG equations for dynamics} \label{sec:tdbdg_equations}
%%%%%%%%%%%%%%%%%%%%%%%%%%%%%%%%%%%%%%%%%%%%%%%%%%%%%%%%%%%%%%%%%%%%%%%%%%%%%%%%%%%%%%%%%%%%%%%%%%%%%%%%%
%%%%%%%%%%%%%%%%%%%%%%%%%%%%%%%%%%%%%%%%%%%%%%%%%%%%%%%%%%%%%%%%%%%%%%%%%%%%%%%%%%%%%%%%%%%%%%%%%%%%%%%%%
%%%%%%%%%%%%%%%%%%%%%%%%%%%%%%%%%%%%%%%%%%%%%%%%%%%%%%%%%%%%%%%%%%%%%%%%%%%%%%%%%%%%%%%%%%%%%%%%%%%%%%%%%

Starting from the self-consistent stationary solution obtained in Sec.~\ref{sec:bdg_equations}, we compute the subsequent dynamics by solving the tdBdG equations~\cite{challisBraggScatteringCooper2007,scottDynamicsDarkSolitons2011a,zouJosephsonOscillationsSelfTrapping2014}. To remove the dc motion of the  potential barrier due to the dc component, we introduce the coordinate transformation $x_{\rm mov }\equiv x-v_{\rm DC}t$, where $x_{\rm mov}$ represents the coordinate in the frame comoving with the dc component of the potential barrier. For simplicity, we denote $x_{\rm mov}$ by $x$ hereafter. The tdBdG equation in this frame is given by
\begin{align}
  i\hbar \frac{\partial}{\partial t} 
  \begin{pmatrix}
    u_{\nu}(x,t) \\
    v_{\nu}(x,t)
  \end{pmatrix}
  =
  \begin{pmatrix}
    h(x,t) & \Delta(x,t) \\
    \Delta^*(x,t) & -h^*(x,t)
  \end{pmatrix}
  \begin{pmatrix}
    u_{\nu}(x,t) \\
    v_{\nu}(x,t)
  \end{pmatrix},
  \label{eq:tdBdG_1D}
\end{align}
where $h(x,t)=-\hbar^2 (\partial^2/ \partial x^2 - k_y^2 - k_z^2)/(2m) + i\hbar v_{\mathrm{DC}}\partial/\partial x + V(x,t) - \mu$ and the order parameter $\Delta(x,t)$ is calculated as 
\begin{align}
  \Delta(x,t) = -\frac{g}{L_{\perp}^2} \sum_{k_y,k_z}\sum_{0\le E_\nu<E_{\mathrm{cut}}} u_{\nu}(x,t) v_{\nu}^*(x,t), \label{eq:tdgap_eq_1D}
\end{align}
self-consistently at each time step. Here, we introduce the term $i\hbar v_{\mathrm{DC}} \partial/\partial x$ in the single-particle Hamiltonian $h(x,t)$, which corresponds to the transformation to a frame moving with the dc component of the barrier velocity, $v_{\mathrm{DC}}$. When evaluating the potential $V(x,t)$, we use $x_{\rm b}(t)=x_0+x_{{\rm AC}}\sin(\omega t)$ instead of Eq.~(\ref{eq:barrier_position}). This transformation allows us to treat the dc component of the potential barrier as static, thereby simplifying the calculation. 
To obtain the dynamics of the particle number density, we calculate
\begin{align}
  n(x,t) = \frac{2}{L_{\perp}^2} \sum_{k_y,k_z} \sum_{E_{\nu} \ge 0} |v_{\nu}(x,t)|^2. \label{eq:tdparticle_number_1D}
\end{align}
Using $n(x,t)$, we evaluate the chemical-potential difference $\Delta \mu(t)$ as explained in Sec.~\ref{sec:potential_barrier}. We note that the charging energy $E_{\mathrm{C}}$ in Eq.~(\ref{eq:chemical_potential_difference}) is fixed at its value evaluated at $t=0$. 

We solve Eqs.~\eqref{eq:tdBdG_1D} and \eqref{eq:tdgap_eq_1D} self-consistently using the split-step Fourier method~\cite{reichlCoreFillingSnaking2017a,tokimotoExcitationHiggsMode2019b,wangCollisionsMajoranaZero2023a} explained in Appendix~\ref{sec:split_step_fourier}.

%%%%%%%%%%%%%%%%%%%%%%%%%%%%%%%%%%%%%%%%%%%%%%%%%%%%%%%%%%%%%%%%%%%%%%%%%%%%%%%%%%%%%%%%%%%%%%%%%%%%%
%%%%%%%%%%%%%%%%%%%%%%%%%%%%%%%%%%%%%%%%%%%%%%%%%%%%%%%%%%%%%%%%%%%%%%%%%%%%%%%%%%%%%%%%%%%%%%%%%%%%%
%%%%%%%%%%%%%%%%%%%%%%%%%%%%%%%%%%%%%%%%%%%%%%%%%%%%%%%%%%%%%%%%%%%%%%%%%%%%%%%%%%%%%%%%%%%%%%%%%%%%%
\section{Results} \label{sec:results}
%%%%%%%%%%%%%%%%%%%%%%%%%%%%%%%%%%%%%%%%%%%%%%%%%%%%%%%%%%%%%%%%%%%%%%%%%%%%%%%%%%%%%%%%%%%%%%%%%%%%%
%%%%%%%%%%%%%%%%%%%%%%%%%%%%%%%%%%%%%%%%%%%%%%%%%%%%%%%%%%%%%%%%%%%%%%%%%%%%%%%%%%%%%%%%%%%%%%%%%%%%%
%%%%%%%%%%%%%%%%%%%%%%%%%%%%%%%%%%%%%%%%%%%%%%%%%%%%%%%%%%%%%%%%%%%%%%%%%%%%%%%%%%%%%%%%%%%%%%%%%%%%%

Throughout this paper, we set the system size along the ring direction to $L_x = 256 k_{\mathrm{F}}^{-1}$ and the system sizes in the transverse directions to $L_{\perp} = 15.2 k_{\mathrm{F}}^{-1}$. For the numerical simulation, the $x$ direction is discretized into $N_x = 512$ grid points, and the transverse wave numbers are restricted to $-14 \le \alpha_y, \alpha_z \le 15$. We set the total particle number to $N = 2000$ and the cutoff energy to $E_{\mathrm{cut}} = 50 E_{\mathrm{F}}$. The height and width of the potential barrier are set to $V_0 = 0.5 E_{\mathrm{F}}$ and $w = k_{\mathrm{F}}^{-1}$, respectively.  We explore different interaction strengths across the BCS-BEC crossover by tuning the dimensionless parameter $(k_{\mathrm{F}}a)^{-1}$.

%%%%%%%%%%%%%%%%%%%%%%%%%%%%%%%%%%%%%%%%%%%%%%%%%%%%%%%%%%%%%%%%%%%%%%%%%%%%%%%%%%%%%%%%%%%%%%%%%%%%%
%%%%%%%%%%%%%%%%%%%%%%%%%%%%%%%%%%%%%%%%%%%%%%%%%%%%%%%%%%%%%%%%%%%%%%%%%%%%%%%%%%%%%%%%%%%%%%%%%%%%%
%%%%%%%%%%%%%%%%%%%%%%%%%%%%%%%%%%%%%%%%%%%%%%%%%%%%%%%%%%%%%%%%%%%%%%%%%%%%%%%%%%%%%%%%%%%%%%%%%%%%%
\subsection{dc driving} \label{sec:dc_driving}
%%%%%%%%%%%%%%%%%%%%%%%%%%%%%%%%%%%%%%%%%%%%%%%%%%%%%%%%%%%%%%%%%%%%%%%%%%%%%%%%%%%%%%%%%%%%%%%%%%%%%
%%%%%%%%%%%%%%%%%%%%%%%%%%%%%%%%%%%%%%%%%%%%%%%%%%%%%%%%%%%%%%%%%%%%%%%%%%%%%%%%%%%%%%%%%%%%%%%%%%%%%
%%%%%%%%%%%%%%%%%%%%%%%%%%%%%%%%%%%%%%%%%%%%%%%%%%%%%%%%%%%%%%%%%%%%%%%%%%%%%%%%%%%%%%%%%%%%%%%%%%%%%

Here, we present our results with pure dc driving, i.e., $x_{\mathrm{AC}} = 0$ in Eq.~\eqref{eq:barrier_position}, as a baseline for the subsequent case with both dc and ac driving. In this case, we calculate the dynamics with a constant barrier displacement of $\Delta x = 50k_{\mathrm{F}}^{-1}$ for all $v_{\mathrm{DC}}$ except for the $v_{\mathrm{DC}} = 0$ case. For numerical efficiency, we set the total number of time steps to $N_t = 2.5\times 10^5$ for all $v_{\mathrm{DC}}$, and the time interval for the numerical integration of the tdBdG equations is determined by $\Delta t = \Delta x/(N_t v_{\mathrm{DC}})$, which ensures that the barrier moves the same distance $\Delta x$ for all $v_{\mathrm{DC}}$ at the end of the simulation time $t_{\mathrm{f}} = N_t \Delta t$.

To extract quantitative information on the observed $ v_{\mathrm{DC}}-\Delta \mu $ characteristics, we use the resistively and capacitively shunted junction (RCSJ) model commonly used to describe superconducting JJs~\cite{baronePhysicsApplicationsJosephson1982, tinkhamIntroductionSuperconductivity2004, stewartCURRENTVOLTAGECHARACTERISTICSJOSEPHSON1968, mccumberEffectAcImpedance1968}. This phenomenological model describes the dynamics of the phase difference $\phi$ across the junction as 
\begin{align}
  I_{\mathrm{ext}} = I_{\mathrm{c}} \sin \phi + \hbar G \dot{\phi} + \frac{\hbar C}{2} \ddot{\phi}, \label{eq:RCSJ}
\end{align}
where $I_{\mathrm{ext}}$ is the external bias current, $I_{\mathrm{c}}$ is the critical current, $G$ is the conductance, and $C = E_{\mathrm{C}}^{-1} $ is the capacitance of the junction. In the dc driving case, the external current bias is given by $I_{\mathrm{ext}} = I_{\mathrm{DC}} = N v_{\mathrm{DC}}/L_x$. We define the phase difference as $\phi=\theta_{\mathrm{L}}-\theta_{\mathrm{R}}$, where $\theta_{\mathrm{L}}$ and $\theta_{\mathrm{R}}$ are the phases of the order parameter in the left and right reservoirs, respectively. The phase difference is related to the chemical-potential difference $\Delta\mu=\mu_{\mathrm{R}}-\mu_{\mathrm{L}}$ by the Josephson--Anderson relation $2\Delta\mu=\hbar\dot{\phi}$, where the factor of 2 arises because $\Delta\mu$ is defined per single fermion rather than per Cooper pair. In the RCSJ model, the Stewart--McCumber parameter $\beta_{\mathrm{c}} = I_{\mathrm{c}} C/(2\hbar G^2)$ characterizes the damping of the junction, with $\beta_{\mathrm{c}} \ll 1$ corresponding to an overdamped regime and $\beta_{\mathrm{c}} \gg 1$ to an underdamped regime. In the overdamped regime, the solution of Eq.~\eqref{eq:RCSJ} is given by 
\begin{align}
  \Delta \mu = (2G)^{-1}\sqrt{I_{\mathrm{DC}}^2 - I_{\mathrm{c}}^2}, \label{eq:RCSJ_solution_overdamped}
\end{align}
for $I_{\mathrm{ext}} > I_{\mathrm{c}}$ and $\Delta \mu = 0$ for $I_{\mathrm{ext}} < I_{\mathrm{c}}$. Following previous studies on superfluid Josephson junctions~\cite{singhShapiroStepsDriven2024b,kwonStronglyCorrelatedSuperfluid2020, delpaceShapiroStepsStronglyinteracting2025a}, we fit the numerical data of $v_{\mathrm{DC}}-\Delta \mu$ characteristics with the overdamped solution to extract the junction-dependent critical driving velocity $v_{\mathrm{c}}$ and the conductance $G$ \footnote{The overdamped solution $\Delta \mu = (2G)^{-1}(I_{\mathrm{DC}}^2 - I_{\mathrm{c}}^2)^{1/2}$ is valid only in the overdamped regime ($\beta_{\mathrm{c}} \ll 1$). However, as shown in Fig.~\ref{fig:dc_fit_params_combined}\textcolor{refcolor}{(c)}, the extracted $\beta_{\mathrm{c}}$ remains greater than unity across the entire crossover, indicating that the junction is in the underdamped regime. Following previous studies~\cite{kwonStronglyCorrelatedSuperfluid2020,singhShapiroStepsDriven2024b,delpaceShapiroStepsStronglyinteracting2025a}, we nonetheless apply this fitting procedure to extract $v_\mathrm{c}$ and $G$ as phenomenological measures, expecting that the qualitative trends across the crossover remain informative. }. Here, $v_{\mathrm{c}}$ is defined operationally as the threshold in the imposed dc barrier velocity for the onset of a finite dissipative response. It is a property of the junction for the present barrier parameters and should not be identified with the Landau critical velocity $v_{\mathrm{L}}^{\mathrm{bulk}}$ of a homogeneous superfluid.

\begin{figure}[htbp]
  \centering
  \includegraphics[width=1.0\linewidth]{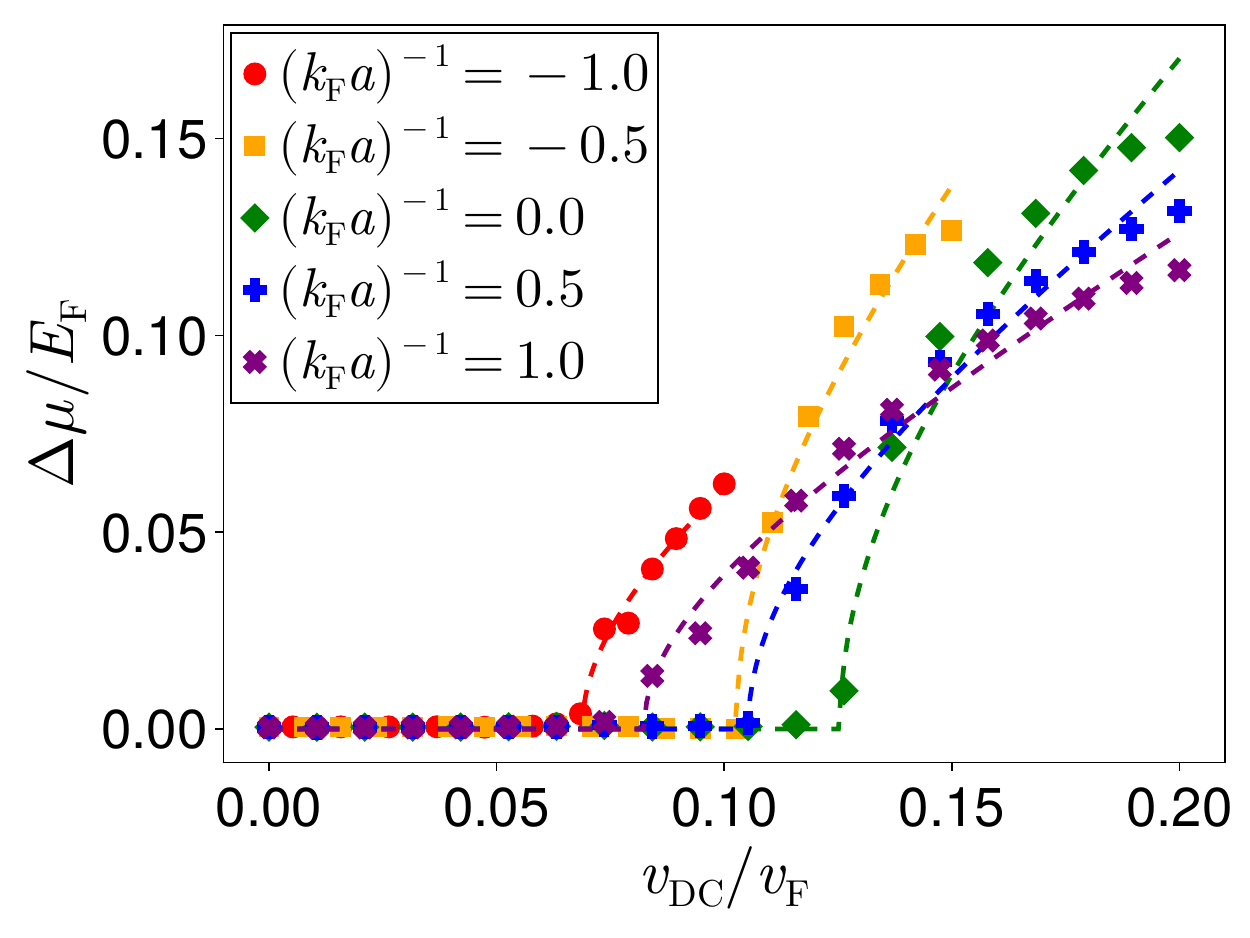}
  \caption{Relationship between the dc barrier velocity $v_{\mathrm{DC}}$ and the chemical-potential difference $\Delta \mu(t_{\mathrm{f}})$ across the BCS-BEC crossover. The symbols show the tdBdG results, and the dashed lines show fits to the overdamped RCSJ expression in Eq.~\eqref{eq:RCSJ_solution_overdamped}, from which the junction-dependent critical driving velocity $v_{\mathrm{c}}$ and the conductance $G$ are extracted.}
  \label{fig:dc_ka_inv}
\end{figure}

Figure~\ref{fig:dc_ka_inv} shows the relationship between the dc barrier velocity $v_{\mathrm{DC}}$ normalized by the Fermi velocity $v_{\mathrm{F}} = \hbar k_{\mathrm{F}}/m$ and the chemical-potential difference $\Delta \mu(t_{\mathrm{f}})$ calculated from a linear fit (see Sec.~\ref{sec:potential_barrier}) for various coupling strengths $(k_{\mathrm{F}}a)^{-1}$. We observe a junction-dependent critical driving velocity $v_{\mathrm{c}}$ below which $ \Delta \mu $ remains zero, indicating a superflow. Above $v_{\mathrm{c}}$, $ \Delta \mu $ increases with $v_{\mathrm{DC}}$, reflecting the energy loss at the junction.

\begin{figure}[htbp]
  \centering
  \includegraphics[width=1.0\linewidth]{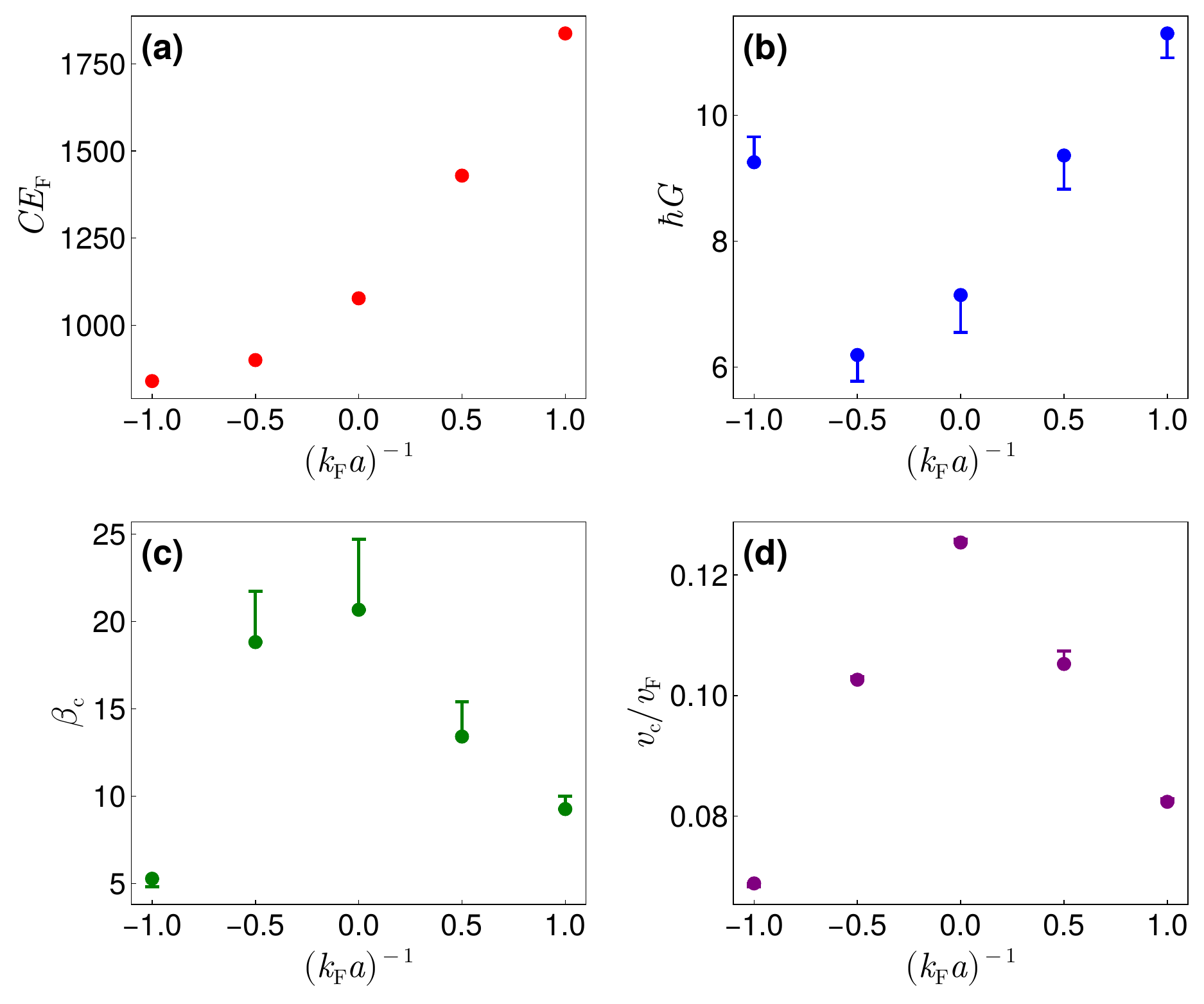}
  \caption{(a) Capacitance $C$, (b) conductance $G$, (c) Stewart--McCumber parameter $\beta_{\mathrm{c}}$, and (d) junction-dependent critical driving velocity $v_{\mathrm{c}}$  as a function of the interaction strength $(k_{\mathrm{F}}a)^{-1}$. $G$ and $v_{\mathrm{c}}$ are extracted from the fits shown in Fig.~\ref{fig:dc_ka_inv}, while $\beta_{\mathrm{c}}$ is calculated from the corresponding junction parameters. The dimensionless capacitance is given by $C E_{\mathrm{F}}=E_{\mathrm{F}}/E_{\mathrm{C}}$  (see Appendix~\ref{sec:charging_energy} for details). The symbols in panels (b)--(d) represent the parameters extracted using the full $v_{\mathrm{DC}}$ data range. The vertical bars indicate the minimum and maximum values obtained by independently excluding up to two data points from each end of the fitting range. These bars represent fitting-range sensitivity rather than statistical confidence intervals. The charging energy $E_{\mathrm{C}}$ is held fixed in this analysis.}
  \label{fig:dc_fit_params_combined}
\end{figure}

\begin{figure*}[t]
  \centering
  \includegraphics[width=1.0\linewidth]{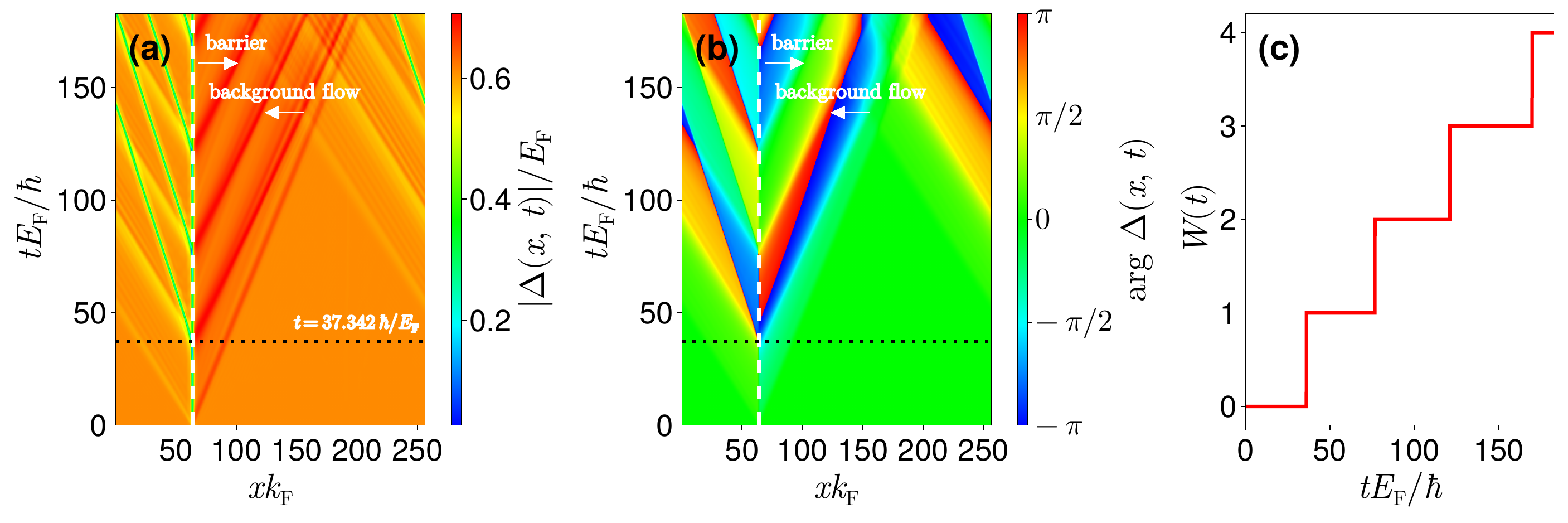}
  \caption{Dynamics of the order parameter $\Delta(x,t)$ in the UFG regime $[(k_{\mathrm{F}}a)^{-1}=0]$ at $v_{\mathrm{DC}} = 0.137 v_{\mathrm{F}}$, which is above the  junction-dependent critical driving velocity $v_{\mathrm{c}}$ (i.e., in the voltage regime). (a) Amplitude $|\Delta(x,t)|$ and (b) phase $\arg\Delta(x,t)$ of the order parameter as functions of position $x$ and time $t$. Note that the potential barrier appears stationary in these plots because the dynamics are calculated in the frame comoving with the dc velocity $v_{\mathrm{DC}}$. The vertical dashed line marks the barrier position. The arrows indicate the direction of the barrier motion in the laboratory frame and the background flow in the frame comoving with the dc component of the barrier. The horizontal dotted line in panels (a) and (b) marks $t=37.342~\hbar/E_{\mathrm{F}}$, at which the spatial profiles shown in Fig.~\ref{fig:soliton_profiles} are evaluated. The nearly vertical depletion near the barrier corresponds to the suppression of the order parameter by the barrier, while the slanted depletion structures propagating away from it correspond to emitted solitons, whose local amplitude and phase structures are analyzed in Fig.~\ref{fig:soliton_profiles}. (c) Time evolution of the winding number $W(t)$ defined in Eq.~\eqref{eq:winding_number}. }
  \label{fig:dc_order_parameter_ufg_voltage}
\end{figure*}

\begin{figure}[t]
    \centering
    \includegraphics[width=1.0\linewidth]{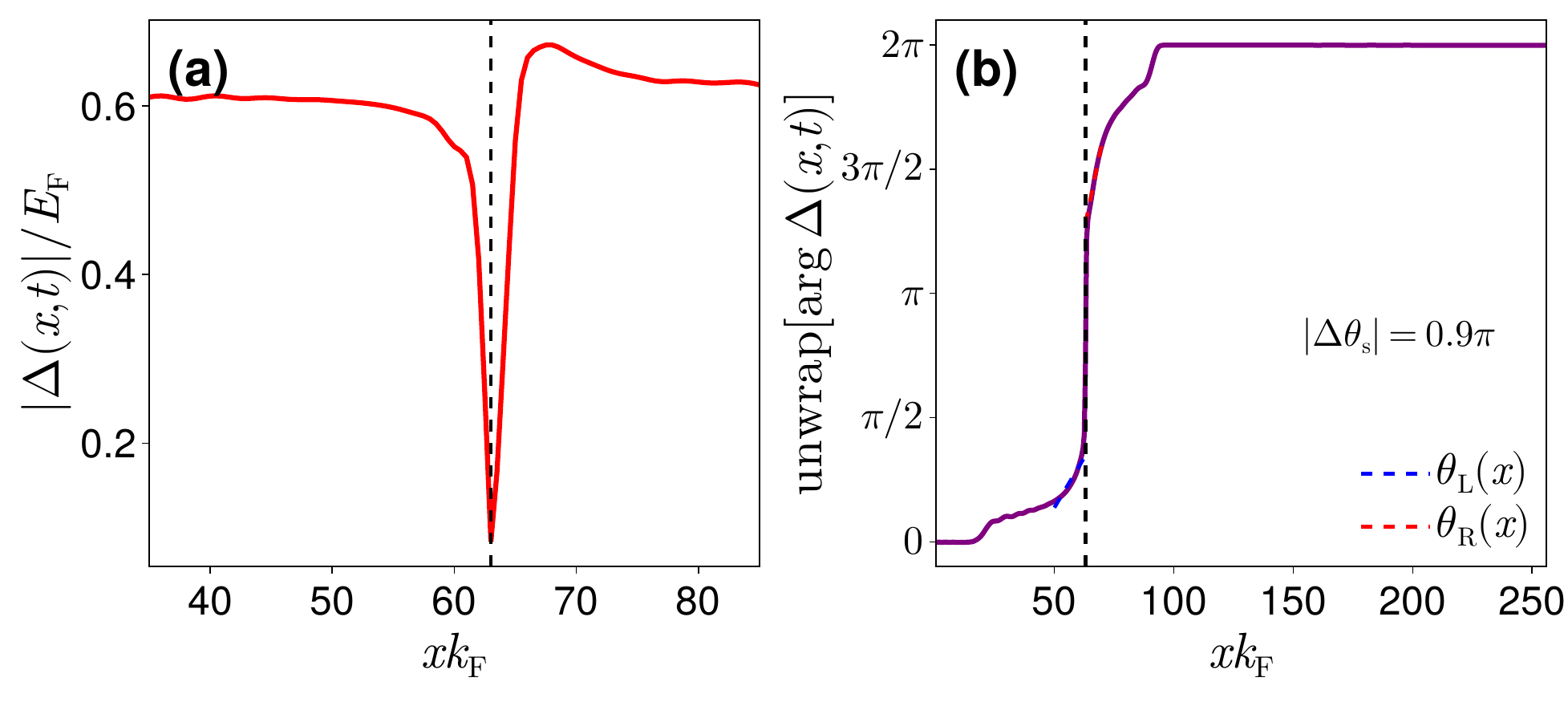}
    \caption{Spatial profiles of the order parameter at $t=37.342~\hbar/E_{\mathrm{F}}$, shortly after a winding number transition from $W=0$ to $W=1$ corresponding to the dynamics shown in Fig.~\ref{fig:dc_order_parameter_ufg_voltage}. The vertical dashed line marks the center $x_{\mathrm{s}}$ of the localized depletion, defined as the position of the minimum of $|\Delta(x,t)|$, in both panels. (a) Enlarged view of the order-parameter amplitude
    $|\Delta(x,t)|/E_{\mathrm F}$ around the localized depletion. (b) Unwrapped phase profile of the order parameter over the full ring. The smooth phase backgrounds on the left and right sides of the depletion are fitted separately by linear functions, denoted by $\theta_{\mathrm{L}}(x)$ and $\theta_{\mathrm{R}}(x)$, respectively. The red and blue dashed lines indicate these linear fits, extrapolated to $x_{\mathrm{s}}$. Their difference at $x_{\mathrm{s}}$, $|\Delta \theta_{\mathrm{s}}| = |\theta_{\mathrm{R}}(x_{\mathrm{s}}) - \theta_{\mathrm{L}}(x_{\mathrm{s}})|$, gives a local phase jump of approximately $ 0.9 \pi $.}
    \label{fig:soliton_profiles}
\end{figure}

The conductance $G$, the resulting Stewart--McCumber parameter $\beta_{\mathrm c}$, and the junction-dependent critical driving velocity $v_{\mathrm c}$ are shown in Figs.~\ref{fig:dc_fit_params_combined}\textcolor{refcolor}{(b)}--\ref{fig:dc_fit_params_combined}\textcolor{refcolor}{(d)}, respectively. We find that Eq.~\eqref{eq:RCSJ_solution_overdamped} reproduces the numerical results well in the dissipative regime above the critical velocity.  The conductance $G$ exhibits a minimum around $(k_{\mathrm{F}}a)^{-1} = -0.5$ and increases toward both the BCS and BEC limits. 
Within the RCSJ framework, the fitted conductance $G$ provides an effective measure of dissipative processes, such as quasiparticle excitations and phonon emission, that accompany the coherent Josephson current. The minimum of $G$ around $(k_{\mathrm{F}}a)^{-1}\simeq -0.5$ therefore suggests that dissipative effects are relatively suppressed in the crossover regime, consistent with the enhanced robustness of superfluid transport near unitarity.
The Stewart--McCumber parameter $\beta_{\mathrm{c}}$ remains well above unity across the entire crossover, suggesting that the junction operates in the underdamped regime throughout. The critical velocity $v_\mathrm{c}$ exhibits a nonmonotonic dependence on interaction strength, peaking in the UFG regime and decreasing toward both the BCS and BEC limits.
We note that this $v_{\mathrm{c}}$ is not the Landau critical velocity $v_{\mathrm{L}}^{\mathrm{bulk}}$ of a homogeneous Fermi superfluid. The latter is determined from the excitation spectrum of the uniform system, whereas the present $v_{\mathrm{c}}$ characterizes the onset of a dissipative response in the presence of a finite barrier. The barrier locally suppresses the density and thereby enhances the local flow velocity, so that a phase-slip instability can be reached even when the imposed barrier velocity $v_{\mathrm{DC}}$ is substantially smaller than $v_{\mathrm{L}}^{\mathrm{bulk}}$.
These results are in good agreement with previous studies on superfluid Josephson junctions~\cite{singhShapiroStepsDriven2024b,spuntarelliJosephsonEffectThroughout2007,spuntarelliSolutionBogoliubovGennes2010a,zouJosephsonOscillationsSelfTrapping2014,valtolinaJosephsonEffectFermionic2015a,kwonStronglyCorrelatedSuperfluid2020, zaccantiCriticalJosephsonCurrent2019}. To assess the sensitivity of the extracted RCSJ parameters to the fitting range, we repeated the fits by independently excluding zero, one, or two data points from each end of the $v_{\mathrm{DC}}$ data range. The vertical bars in Figs.~\ref{fig:dc_fit_params_combined}\textcolor{refcolor}{(b)}--\ref{fig:dc_fit_params_combined}\textcolor{refcolor}{(d)} indicate the minimum and maximum values obtained from the resulting nine fitting windows. Although some fitting-range dependence is observed, particularly for $\beta_{\mathrm{c}}$, the qualitative interaction dependence remains unchanged, and $\beta_{\mathrm{c}}$ remains well above unity throughout the crossover.

The dynamics of the amplitude and phase of the order parameter $\Delta(x,t)$ in the UFG regime at $v_{\mathrm{DC}} = 0.137 v_{\mathrm{F}}$, which is above the junction-dependent critical driving velocity $v_{\mathrm{c}}$ (i.e., in the voltage regime), are shown in Fig.~\ref{fig:dc_order_parameter_ufg_voltage}. We observe that, in the voltage regime ($v_{\mathrm{DC}} > v_{\mathrm{c}}$), the depletion waves in the order-parameter amplitude $|\Delta (x,t)|$ are generated at the barrier  
and propagate in the direction opposite to the barrier motion. 
In the frame comoving with the dc component of the barrier, the background superfluid flows toward the negative $x$ direction relative to the barrier. Therefore, the propagation of the depletion waves toward negative $x$ corresponds to downstream emission associated with this background flow, rather than being caused by the coordinate transformation itself. Similar backward emission of depletion waves relative to the barrier motion has also been reported in previous studies of driven atomic Josephson junctions~\cite{bernhartObservationShapiroSteps2025a,delpaceShapiroStepsStronglyinteracting2025a}.
This process is repeated periodically, which leads to an increase in the chemical-potential difference $\Delta \mu$ between the two reservoirs. To characterize this process, we calculate the winding number $W(t)$ defined as 
\begin{align}
  W(t) = \frac{1}{2\pi} \int_0^{L_x} dx~ \frac{\partial}{\partial x} \arg \Delta(x,t). \label{eq:winding_number}
\end{align}
Numerically, $W(t)$ is evaluated by summing the phase differences between neighboring grid points after phase unwrapping to remove the $2\pi$ ambiguity. As shown in Fig.~\ref{fig:dc_order_parameter_ufg_voltage}\textcolor{refcolor}{(c)}, $W(t)$ increases by one whenever a depletion wave is generated at the barrier, indicating a global $2\pi$ phase-slip event of the order parameter around the ring.

To examine the local phase structure of an emitted depletion wave, we show in Fig.~\ref{fig:soliton_profiles} spatial profiles of the order-parameter amplitude and the unwrapped phase shortly after the first winding-number transition. To remove the artificial $2\pi$ discontinuities associated with the principal-value representation, the phase is unwrapped along the ring direction before the local structure is analyzed.
As shown in Fig.~\ref{fig:soliton_profiles}(a), the propagating depletion exhibits a pronounced localized minimum of $|\Delta|$. To extract the phase jump associated with this structure, we fit the smooth phase backgrounds on the left and right sides of the depletion separately by linear functions, $\theta_{\mathrm{L}}(x)$ and $\theta_{\mathrm{R}}(x)$, and extrapolate them to the depletion center $x_{\mathrm{s}}$. This gives $|\Delta \theta_{\mathrm{s}} | = |\theta_{\mathrm{R}}(x_{\mathrm{s}}) - \theta_{\mathrm{L}}(x_{\mathrm{s}})| \simeq 0.9\pi$. 
The localized amplitude minimum together with this approximately $\pi$ phase jump supports the identification of the emitted depletion wave as a soliton.
 
We note that the stability of these solitons is specific to the effective one-dimensional geometry considered here, in which transverse spatial inhomogeneity and hence the snake instability are excluded. In realistic higher-dimensional geometries, the corresponding phase-slip process may instead involve vortex-antivortex pairs or vortex rings, as found in recent atomic Josephson-junction experiments and their accompanying numerical simulations~\cite{bernhartObservationShapiroSteps2025a,delpaceShapiroStepsStronglyinteracting2025a}.

In contrast, in the superflow regime ($v_{\mathrm{DC}} < v_{\mathrm{c}}$), no depletion wave is generated, and the phase profile remains smooth without any phase slip, which results in $\Delta \mu = 0$. In the present effective one-dimensional geometry, the transition to the voltage state is driven by the periodic emission of solitons. This phase-slip-mediated mechanism of dissipative transport is consistent with earlier studies~\cite{hakimNonlinearSchrodingerFlow1997a,burchiantiConnectingDissipationPhase2018,xhaniCriticalTransportVortex2020}.

\begin{figure}[htbp]
  \centering
  \includegraphics[width=1.0\linewidth]{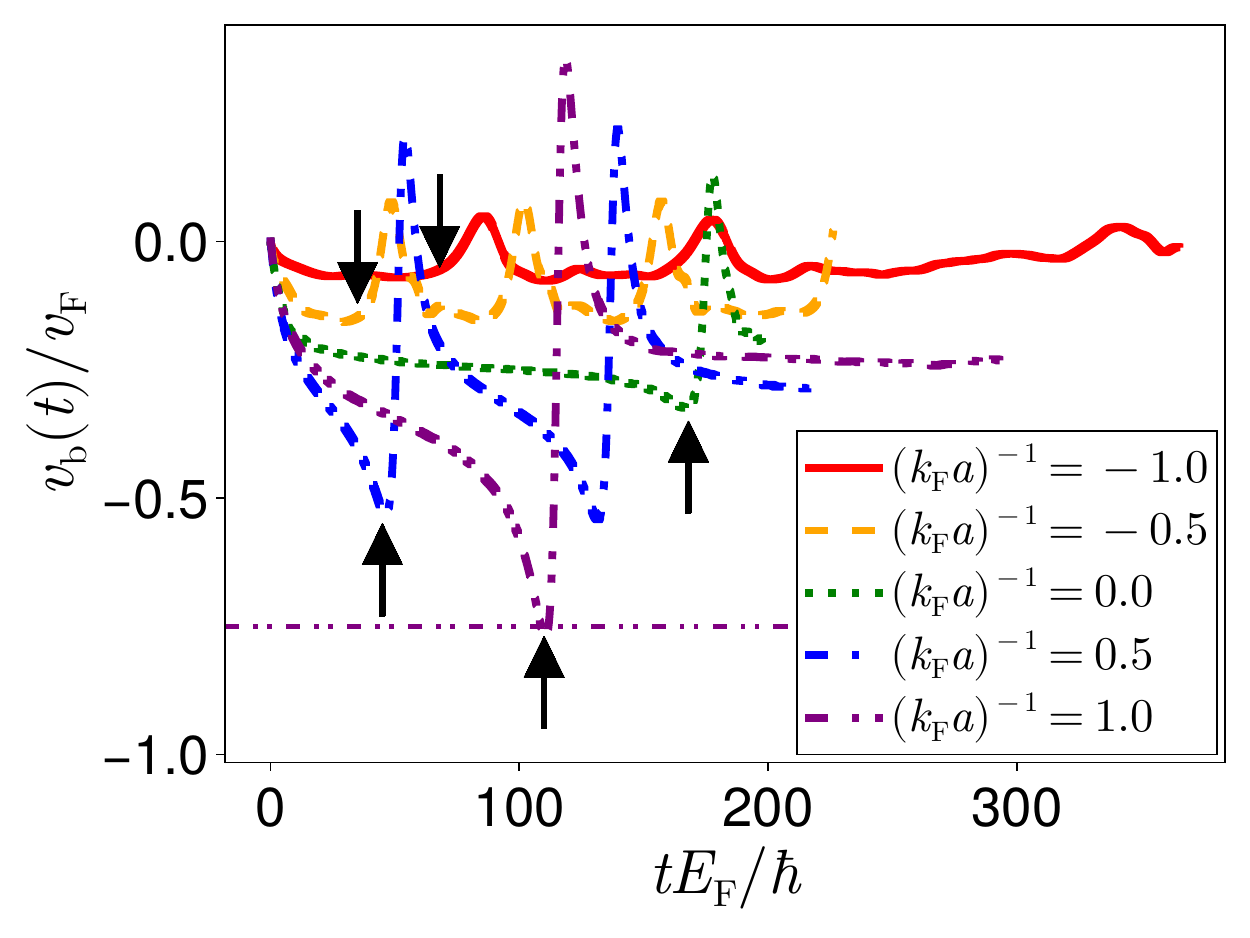}
  \caption{Time evolution of the local velocity $v_{\mathrm{b}}(t) = j(x_{\mathrm{b}},t)/n(x_{\mathrm{b}},t) $ at the barrier position under dc driving. Each curve corresponds to the smallest dc velocity $v_{\mathrm{DC}}$ at which soliton emission is first observed for each interaction strength $(k_{\mathrm{F}}a)^{-1}$, with $((k_{\mathrm{F}}a)^{-1}, v_{\mathrm{DC}}/v_{\mathrm{F}}) = (-1.0, 0.07),\allowbreak (-0.5, 0.12),\allowbreak (0.0, 0.13),\allowbreak (0.5,0.12),\allowbreak (1.0, 0.08) $. Arrows indicate the approximate times of the first soliton-emission events in each case, identified by abrupt changes in $v_{\mathrm{b}}(t)$, typically from negative to positive values. The local critical velocity $v_{\mathrm{lc}}$ is defined as the magnitude of the minimum value of $v_{\mathrm{b}}$ immediately before the first soliton-emission event. The horizontal dash-dot-dot line indicates $-v_{\mathrm{lc}}/v_{\mathrm{F}}$ for $(k_{\mathrm{F}}a)^{-1}=1.0$, highlighting its significantly larger magnitude compared with the BCS and UFG regimes.}
  \label{fig:local_velocity_ka_inv}
\end{figure}

To gain further microscopic insight into the soliton emission mechanism and its dependence on the interaction strength, we analyze the local superfluid velocity at the barrier position. We define the local current density in the laboratory frame as~\cite{spuntarelliSolutionBogoliubovGennes2010a,wlazlowskiDissipationMechanismsFermionic2023}
\begin{align}
  j(x,t) = \frac{2\hbar}{m L_\perp^2} \sum_{k_y,k_z} \sum_{E_\nu \ge 0} \mathrm{Im}\left[ v_{\nu}(x,t) \frac{\partial}{\partial x} v^*_{\nu}(x,t)\right], \label{eq:current_density}
\end{align}
and the local velocity at the barrier position $x_{\mathrm{b}}(t)$ as
\begin{align}
  v_{\mathrm{b}}(t) = \frac{j(x_{\mathrm{b}},t)}{n(x_{\mathrm{b}},t)} \equiv \frac{j_{\mathrm{b}}(t)}{n_{\mathrm{b}}(t)}, \label{eq:local_velocity}
\end{align}
where $n_{\rm b}(t)$ is the local particle density at the barrier. 
Although the tdBdG equations are expressed in the coordinate frame comoving with the dc component of the barrier, the transformation is performed without a momentum boost. Therefore, the current density defined above is the laboratory-frame current density expressed in the comoving spatial coordinate \footnote{Explicitly, $v_\nu(x,t)=v_\nu^{(\mathrm{L})}(x_{\mathrm{L}},t), \quad x_{\mathrm{L}}=x+v_{\mathrm{DC}}t,$ and $\partial_x=\partial_{x_{\mathrm{L}}}$, which gives $j(x,t)=j^{(\mathrm{L})}(x_{\mathrm{L}},t)$.}.
We distinguish the imposed barrier velocity $v_{\mathrm{DC}}$, the local laboratory-frame flow velocity $v_{\mathrm{b}}(t)$ at the barrier, and the homogeneous bulk critical velocity $v_{\mathrm{L}}^{\mathrm{bulk}}$; the fitted $v_{\mathrm{c}}$ is the threshold value of $v_{\mathrm{DC}}$ for the onset of the dissipative junction response.

Figure~\ref{fig:local_velocity_ka_inv} shows the time evolution of the local velocity $v_{\mathrm{b}}(t)$ at the smallest dc velocity $v_{\mathrm{DC}}$ at which soliton emission is first observed for each interaction regime, corresponding to $ ((k_{\mathrm{F}}a)^{-1}, v_{\mathrm{DC}}/v_{\mathrm{F}}) = (-1.0, 0.07),\allowbreak (-0.5, 0.12),\allowbreak (0.0, 0.13),\allowbreak (0.5,0.12),\allowbreak (1.0, 0.08) $. The approximate times of the first soliton-emission events are indicated by the arrows and coincide with abrupt changes in $v_{\mathrm b}(t)$, typically from negative to positive values. We define the local critical velocity $v_{\mathrm{lc}}$ as the magnitude of the minimum value of $v_{\mathrm b}(t)$ immediately before the first soliton-emission event. As is evident from Fig.~\ref{fig:local_velocity_ka_inv}, $v_{\mathrm{lc}}$ increases monotonically from the BCS to the BEC regime. This trend is in stark contrast to the critical driving velocity $v_{\mathrm{c}}$ shown in Fig.~\ref{fig:dc_fit_params_combined}\textcolor{refcolor}{(d)}, which exhibits a nonmonotonic dependence with a maximum at unitarity. The monotonic increase of $v_{\mathrm{lc}}$ toward the BEC regime reflects the enhanced local density depletion at the barrier: the reduced local density $n_{\rm b}(t)/n_0$ amplifies the local velocity $v_{\mathrm{b}}(t) = j_{\rm b}(t)/n_{\rm b}(t)$ relative to the bulk superfluid velocity, resulting in a larger effective threshold for soliton emission. This enhanced depletion is consistent with the increased compressibility of the system in the BEC regime, as reflected in the larger capacitance $C$ shown in Fig.~\ref{fig:dc_fit_params_combined}\textcolor{refcolor}{(a)}, which makes the local density more susceptible to suppression by the barrier potential. Such a mechanism, where the suppression of the local density by the barrier dictates the onset of instability, is  consistent with the previous work based on the local density approximation~\cite{watanabeCriticalVelocitySuperfluid2009a}.

%%%%%%%%%%%%%%%%%%%%%%%%%%%%%%%%%%%%%%%%%%%%%%%%%%%%%%%%%%%%%%%%%%%%%%%%%%%%%%%%%%%%%%%%%%%%%%%%%%%%%
%%%%%%%%%%%%%%%%%%%%%%%%%%%%%%%%%%%%%%%%%%%%%%%%%%%%%%%%%%%%%%%%%%%%%%%%%%%%%%%%%%%%%%%%%%%%%%%%%%%%%
%%%%%%%%%%%%%%%%%%%%%%%%%%%%%%%%%%%%%%%%%%%%%%%%%%%%%%%%%%%%%%%%%%%%%%%%%%%%%%%%%%%%%%%%%%%%%%%%%%%%%
\subsection{dc and ac driving} \label{sec:ac_driving}
%%%%%%%%%%%%%%%%%%%%%%%%%%%%%%%%%%%%%%%%%%%%%%%%%%%%%%%%%%%%%%%%%%%%%%%%%%%%%%%%%%%%%%%%%%%%%%%%%%%%%
%%%%%%%%%%%%%%%%%%%%%%%%%%%%%%%%%%%%%%%%%%%%%%%%%%%%%%%%%%%%%%%%%%%%%%%%%%%%%%%%%%%%%%%%%%%%%%%%%%%%%
%%%%%%%%%%%%%%%%%%%%%%%%%%%%%%%%%%%%%%%%%%%%%%%%%%%%%%%%%%%%%%%%%%%%%%%%%%%%%%%%%%%%%%%%%%%%%%%%%%%%%

Here, we present our results with both dc and ac driving, i.e., $x_{\mathrm{AC}} \neq 0$. We set the total simulation time $t_{\mathrm{f}}$ to exactly three cycles ($N_{\mathrm{cycle}} = 3$) of the ac modulation. This duration is chosen to capture the synchronized phase-slip dynamics while avoiding substantial system-wide disturbance caused by the recirculation of emitted excitations. The time step for the numerical integration of the tdBdG equations is chosen as $\Delta t = 0.005\hbar/E_{\mathrm{F}}$ for $(k_{\mathrm{F}}a)^{-1} \le 0$, $\Delta t = 0.002 \hbar/E_{\mathrm{F}}$ for $(k_{\mathrm{F}}a)^{-1} = 0.5$, and $\Delta t = 0.001 \hbar/E_{\mathrm{F}}$ for $(k_{\mathrm{F}}a)^{-1}=1$. By systematically varying the dc velocity $v_{\mathrm{DC}}$, the ac modulation amplitude $x_{\mathrm{AC}}$, and the frequency $\omega$, we investigate their respective effects on the transport properties.

\begin{figure}[ht]
  \centering
  \includegraphics[width=1.0\linewidth]{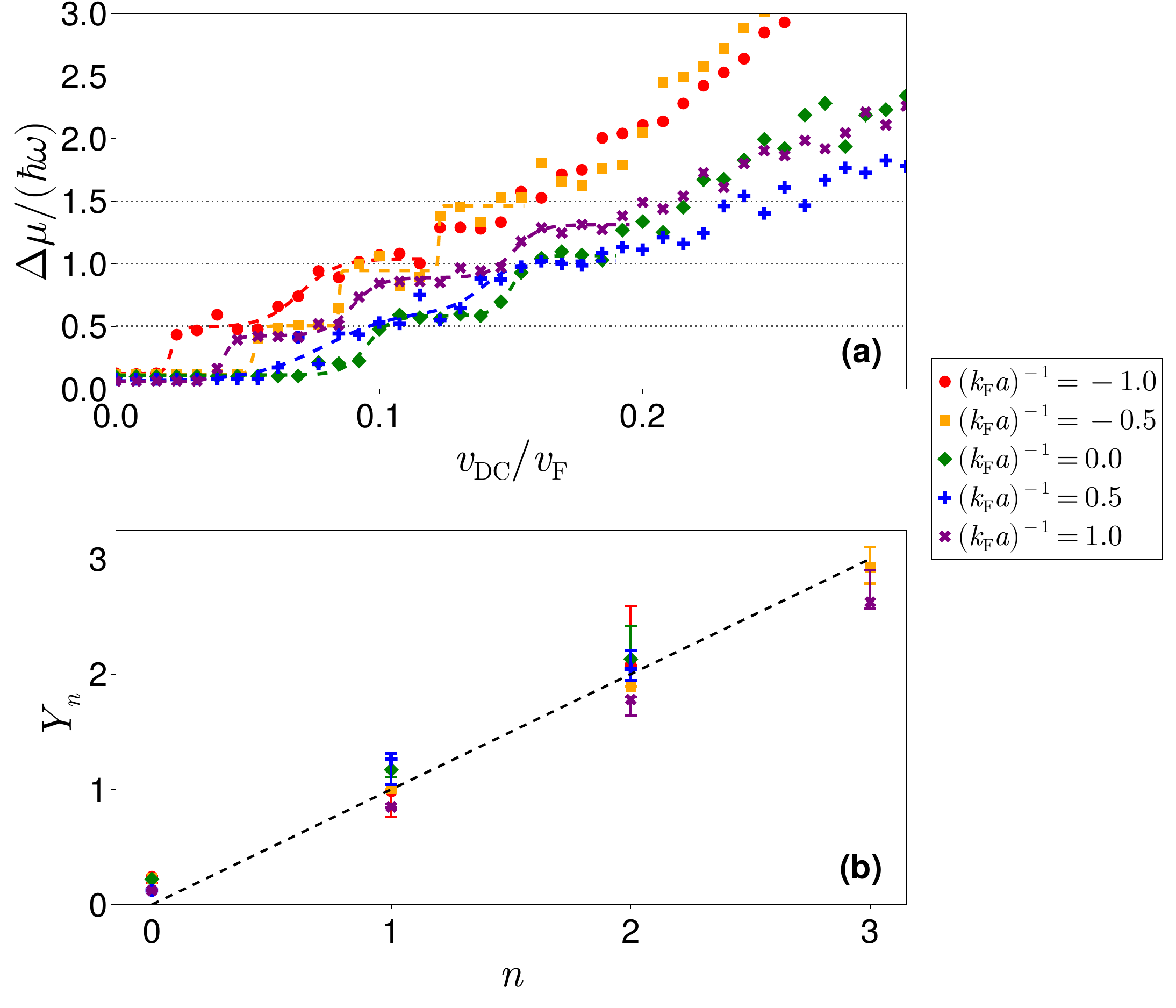}
  \caption{(a) Relationship between the dc barrier velocity $v_{\mathrm{DC}}$ and the chemical-potential difference $\Delta \mu(t_{\mathrm{f}})$ across the BCS-BEC crossover in the presence of ac modulation. Symbols show the tdBdG results and dashed curves show multiple-logistic fits. Black dotted horizontal lines indicate the theoretical quantized values of $\Delta \mu$ at integer multiples of $\hbar \omega/2$. The parameters of the ac modulation are set to $x_{\mathrm{AC}} = 3 k_{\mathrm{F}}^{-1}$ and $\omega = 0.04 E_{\mathrm{F}}/\hbar$. (b) Plateau levels $P_n$ extracted from the fits, plotted as $Y_n \equiv 2P_n/(\hbar\omega)$ versus the step index $n$. The dashed line represents $Y_n = n$. Symbols indicate the values obtained using the nominal fitting window, while vertical bars indicate the minimum and maximum values of $Y_n$ obtained by independently shifting each endpoint of the fitting window by up to two data points, with $M$ kept fixed.}
  \label{fig:shapiro_ka_inv}
\end{figure}

Figure~\ref{fig:shapiro_ka_inv}\textcolor{refcolor}{(a)} shows the relationship between the dc barrier velocity $v_{\mathrm{DC}}$ and the chemical-potential difference $\Delta\mu(t_{\mathrm{f}}) $ calculated from the linear fit (see Sec.~\ref{sec:potential_barrier}) in the presence of ac modulation for varying interaction strengths $(k_{\mathrm{F}}a)^{-1}$. 
The numerical curves exhibit low-order plateau-like structures near integer multiples of $\hbar\omega/2$, although the plateaus are broadened and accompanied by additional structures, particularly at higher biases. To determine whether these plateau-like features are consistent with the expected Shapiro quantization, we quantitatively extract their plateau levels. Motivated by the logistic and sigmoidal fitting procedures used in recent experimental studies of Shapiro steps~\cite{bernhartObservationShapiroSteps2025a,delpaceShapiroStepsStronglyinteracting2025a}, we fit each $v_{\mathrm{DC}}-\Delta \mu$ curve with a sum of logistic functions
\begin{align}
  f(v_{\mathrm{DC}}) = P_0 + \sum_{j=1}^M \frac{A_j}{1 + \exp[-k_j (v_{\mathrm{DC}} - v_j)]}.
\end{align}
Here, $P_0$, $A_j$, $k_j$, and $v_j$ are fitting parameters: $P_0$ is the baseline value of $\Delta\mu$ before the first transition, $A_j$ is the height increment, $k_j$ controls the steepness, and $v_j$ specifies the center of the $j$th transition. For each interaction strength, we select a nominal fitting window encompassing the visually resolved low-order transition region and set $M$ to the number of transitions included in that window. Specifically, we use $M=2,3,2,2,$ and $3$ for $(k_{\mathrm F}a)^{-1}=-1.0,-0.5,0.0,0.5,$ and $1.0$, respectively. To assess the sensitivity to this choice, each endpoint of the nominal fitting window is independently shifted by up to two data points while $M$ is kept fixed. The height parameters $A_j$ are treated as independent fitting parameters, so that the expected Shapiro quantization is not imposed in the fitting procedure. The $n$th plateau level is then obtained as $P_n = P_0 + \sum_{j=1}^n A_j$.
Figure~\ref{fig:shapiro_ka_inv}\textcolor{refcolor}{(b)} shows the extracted plateau levels in terms of $Y_n \equiv 2P_n/(\hbar\omega)$. The low-order plateau levels lie close to the expected relation $Y_n = n$ over the interaction regimes considered. The vertical ranges indicate the minimum and maximum values obtained by varying each end of the fitting window by up to two data points. The agreement is less precise for some broadened and higher-order plateaus, indicating that the ideal quantized staircase is not realized uniformly over the entire bias range.

The plateau levels are relatively insensitive to the interaction strength, consistent with their origin in the phase-locking condition set by the external drive frequency. In contrast, the onset positions, widths, and visibility of the plateaus vary across the crossover, reflecting the interaction dependence of the junction properties. We also emphasize that the numerical characteristics do not form an ideal staircase over the entire bias range. The additional structures and broadening, particularly near plateau boundaries and at higher biases, can arise from the finite observation time and the underdamped transient dynamics, together with discrete soliton and phonon emission events. In the BCS regime, pair breaking at high bias further degrades phase coherence, as discussed below.

 Here, the expected quantization unit $\hbar\omega/2$ is half of the standard Shapiro-step spacing $\hbar\omega$ reported in Refs.~\cite{bernhartObservationShapiroSteps2025a,delpaceShapiroStepsStronglyinteracting2025a}. In the experimental convention used in Ref.~\cite{delpaceShapiroStepsStronglyinteracting2025a}, the reported chemical-potential bias is the one entering the Josephson-Anderson relation for the condensate phase, i.e., the pair chemical-potential difference. In contrast, the chemical potential $\mu$ in the BdG equations [Eq.~\eqref{eq:BdG_1D}] is defined per single fermion rather than per Cooper pair ($\mu_{\mathrm{pair}} \simeq 2\mu$). Consequently, the chemical-potential difference $\Delta\mu$ calculated here is half the value that would appear in the standard Josephson-Anderson relation ($\Delta\mu_{\mathrm{pair}} = \hbar\dot{\phi}$). This is merely a formal choice of definition and does not alter the underlying topological physics of the Shapiro steps.

\begin{figure}[htbp]
  \centering
  \includegraphics[width=1.0\linewidth]{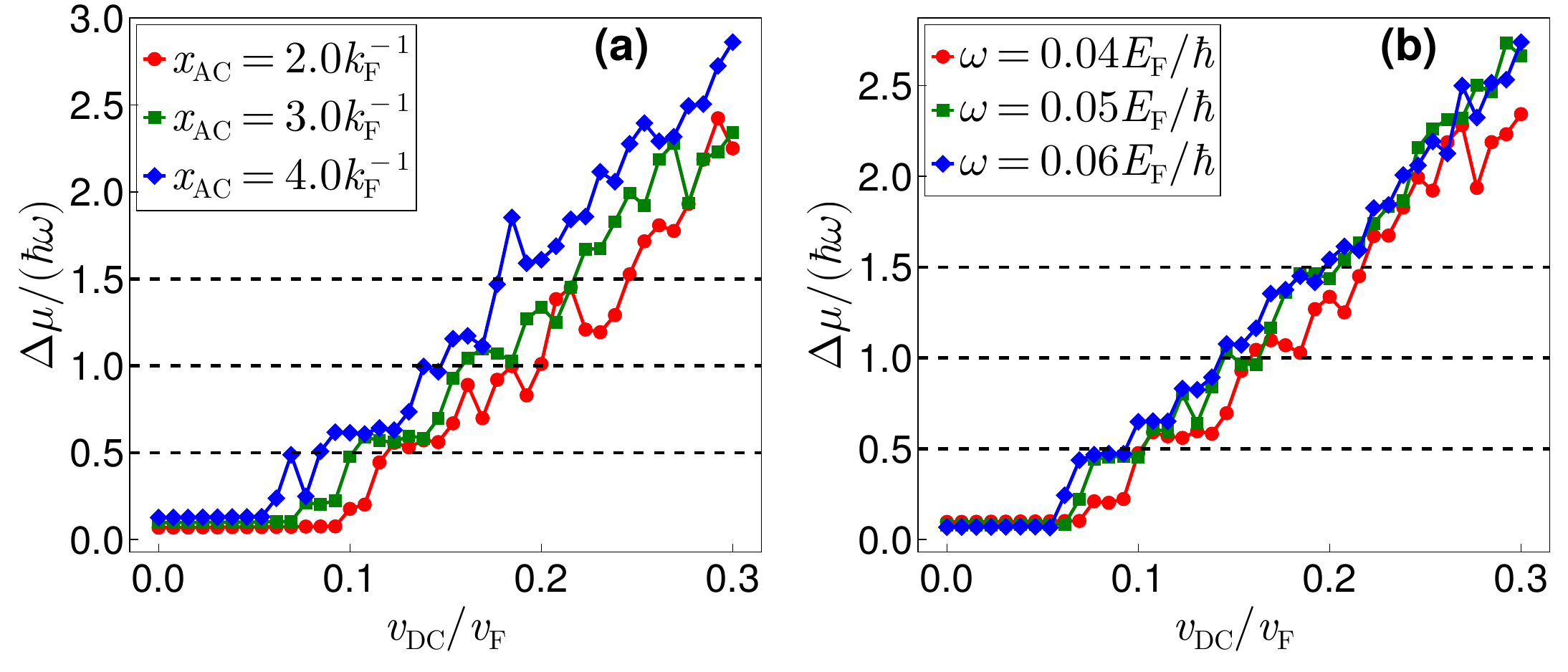}
  \caption{(a) Amplitude dependence and (b) frequency dependence of the Shapiro steps in the UFG regime. The parameters of the ac modulation are set to $\omega = 0.04 E_{\mathrm{F}}/\hbar$ for (a) and $x_{\mathrm{AC}} = 3 k_{\mathrm{F}}^{-1}$ for (b). The vertical axes are normalized by the respective $\hbar \omega$. Black dashed horizontal lines indicate the quantized values of $\Delta \mu$ at integer multiples of $\hbar \omega/2$.}
  \label{fig:shapiro_amp_freq_ufg}
\end{figure}

\begin{figure*}[t]
  \centering
  \includegraphics[width=1.0\linewidth]{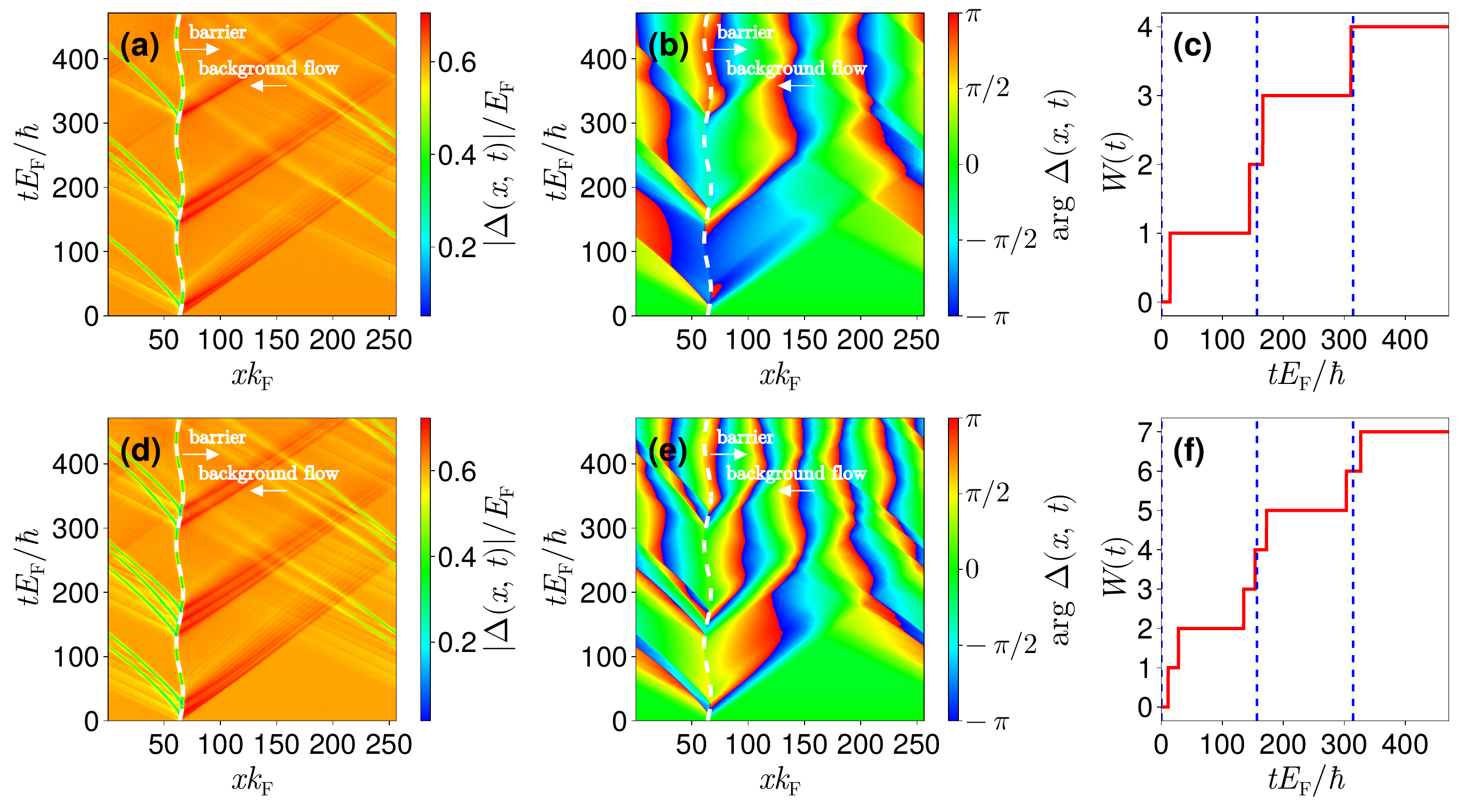}
  \caption{Dynamics of the amplitude and phase of the order parameter $\Delta(x,t)$ in the UFG regime [$(k_{\mathrm{F}}a)^{-1}=0$]. The upper panels (a)--(c) correspond to the first step ($v_{\mathrm{DC}} = 0.123 v_{\mathrm{F}}$), and the lower panels (d)--(f) correspond to the second step ($v_{\mathrm{DC}} = 0.154 v_{\mathrm{F}}$). Panels (a) and (d) show the amplitude $|\Delta(x,t)|$, panels (b) and (e) show the phase $\arg \Delta(x,t)$, and panels (c) and (f) show the winding number $W(t)$ defined in Eq.~\eqref{eq:winding_number}. Blue dashed lines in panels (c) and (f) indicate $t = 2n\pi/\omega~(n=0,1,2)$ when the barrier velocity $\dot{x}_{\mathrm{b}}$ is maximum. Note that the potential barrier appears to move only in an ac manner in these plots because the dynamics are calculated in the frame comoving with the dc velocity $v_{\mathrm{DC}}$. The dashed curves mark the instantaneous barrier position $x_{\mathrm{b}}(t) = x_0 + x_{\mathrm{AC}}\sin (\omega t)$ in the comoving frame. The arrows in panels (a), (b), (d), and (e) indicate the direction of the barrier motion in the laboratory frame and the background flow in the frame comoving with the dc component of the barrier. The oscillatory depletion near the barrier reflects the ac motion of the barrier, while the slanted depletion structures propagating away from it correspond to emitted solitons. The parameters of the ac modulation are set to $x_{\mathrm{AC}} = 3 k_{\mathrm{F}}^{-1}$ and $\omega = 0.04 E_{\mathrm{F}}/\hbar$.}
  \label{fig:amp_phase_ufg}
\end{figure*}

Figure~\ref{fig:shapiro_amp_freq_ufg} shows the amplitude and frequency dependence of the Shapiro steps in the UFG regime. As shown in Fig.~\ref{fig:shapiro_amp_freq_ufg}\textcolor{refcolor}{(a)}, the plateau levels are approximately independent of the ac modulation amplitude $x_{\mathrm{AC}}$. Furthermore, Fig.~\ref{fig:shapiro_amp_freq_ufg}\textcolor{refcolor}{(b)} elucidates the frequency dependence by plotting the chemical-potential difference scaled by the modulation energy, $\Delta \mu / (\hbar \omega)$. The plateaus for three different frequencies roughly coincide at the same values on this scaled axis. This approximate collapse supports the proportionality of the plateau height to the modulation frequency $\omega$.
The weak dependence of the plateau levels on $x_{\mathrm{AC}}$, together with their proportionality to $\hbar\omega$, is consistent with the expected relation $\Delta\mu=n\hbar\omega/2$, with $n=0,1,2,\ldots$, and thus provides additional support for identifying the low-order plateaus as Shapiro steps.

\begin{figure}[htbp]
  \centering
  \includegraphics[width=1.0\linewidth]{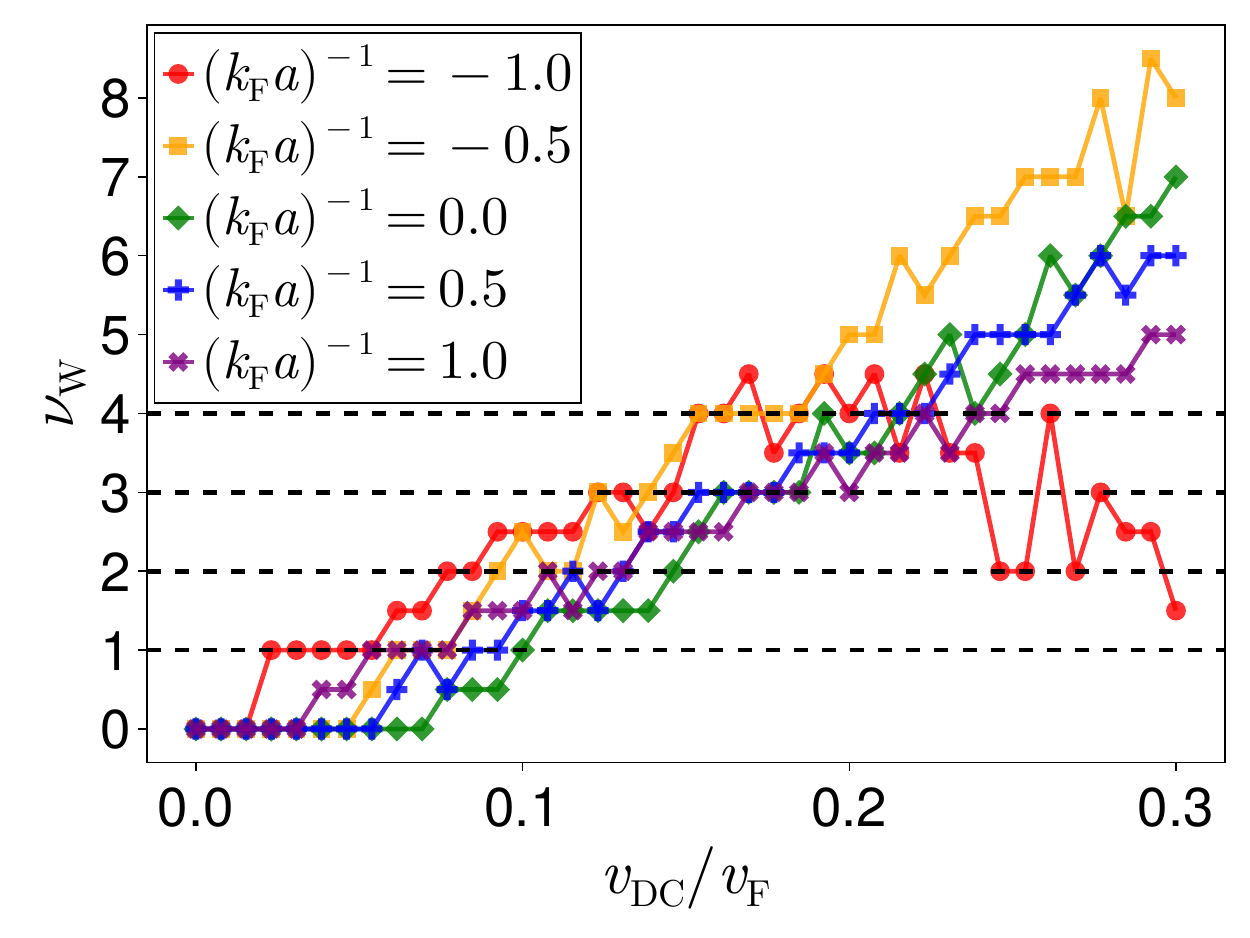}
  %\caption{Winding number $W(t_{\mathrm{f}})$ at the end of the simulation time $t_{\mathrm{f}}$ as a function of the dc barrier velocity $v_{\mathrm{DC}}$ for varying interaction strengths $(k_{\mathrm{F}}a)^{-1}$. The parameters of the ac modulation are set to $x_{\mathrm{AC}} = 3 k_{\mathrm{F}}^{-1}$, $\omega = 0.04 E_{\mathrm{F}}/\hbar$ and $N_{\mathrm{cycle}} = 3$. }
  \caption{Net winding-number increment per drive cycle, $\nu_{\mathrm{W}} = [W(5T/2) - W(T/2)]/2$ as a function of the dc barrier velocity $v_{\mathrm{DC}}$ for varying interaction strengths $(k_{\mathrm{F}}a)^{-1}$, where $T=2\pi/\omega$. The quantity $\nu_{\mathrm{W}}$ is evaluated over the central two cycles of the three-cycle driving protocol. Horizontal dashed lines indicate integer values corresponding to ideal cycle-to-cycle locking. The parameters of the ac modulation are $x_{\mathrm{AC}} = 3k_{\mathrm{F}}^{-1}$ and $\omega = 0.04 E_{\mathrm{F}}/\hbar$. }
  \label{fig:winding_number_ka_inv}
\end{figure}

To connect this macroscopic plateau structure with the microscopic phase dynamics, we next examine the order-parameter dynamics on the first and second plateaus. The dynamics of the order parameter $\Delta(x,t)$ at the first ($v_{\mathrm{DC}} = 0.123 v_{\mathrm{F}}$) and second ($v_{\mathrm{DC}} = 0.154 v_{\mathrm{F}}$) steps in the UFG regime are shown in Fig.~\ref{fig:amp_phase_ufg}. We observe that, similar to the dc driving case, solitons are generated at the barrier and propagate  downstream, i.e., in the direction opposite to the dc barrier motion. In contrast with the dc-driven case, the generation of solitons is synchronized with the ac modulation and underlies the observed low-order plateau response near integer multiples of $\hbar\omega/2$.

To quantify the synchronization of the winding-number dynamics while reducing the influence of phase-slip events near the beginning and end of the finite driving protocol, we define the net winding-number increment per drive cycle as 
\begin{align}
 \nu_{\mathrm{W}} \equiv \frac{W(5T/2) - W(T/2)}{2}, \quad T \equiv \frac{2\pi}{\omega}.
\end{align}
Here, the interval from $T/2$ to $5T/2$ covers the central two driving cycles of the three-cycle protocol. The half-period-shifted boundaries are chosen to avoid counting phase-slip events occurring close to the endpoints $t = 0$ and $3T$, where the instantaneous barrier velocity is maximal. As shown in Fig.~\ref{fig:winding_number_ka_inv}, $\nu_{\mathrm{W}}$ exhibits step-like behavior correlated with the low-order Shapiro-step regions. Since $\nu_{\mathrm{W}}$ is evaluated over only two driving cycles, it can take half-integer values when the numbers of winding-number changes differ by one between the two cycles. Such half-integer values therefore reflect finite-time or transitional
phase-slip dynamics and do not by themselves indicate fractional Shapiro locking.
Nevertheless, the step-like evolution of $\nu_{\mathrm{W}}$ indicates that the winding-number changes, and hence the global phase-slip events, are synchronized with the external modulation at the level resolved by the present finite-time protocol.

Furthermore, it is important to note the behavior in the BCS regime at high dc biases (e.g., $v_{\mathrm{DC}}/v_{\mathrm{F}} \gtrsim 0.2$). In this weak-coupling regime, the pairing gap $\Delta$ is relatively small, leading to a low Landau critical velocity for pair-breaking excitations [$v_{\mathrm{pb}} = \{[(\mu^2 + \Delta^2)^{1/2} - \mu]/m\}^{1/2} \approx 0.1 v_{\mathrm{F}}$ for $\Delta \approx 0.2 E_{\mathrm{F}}$ at $(k_{\mathrm{F}}a)^{-1} = -1.0$]~\cite{combescotCollectiveModeHomogeneous2006}. At the highest dc biases considered here, $v_{\mathrm{DC}}$ is comparable to or exceeds this bulk pair-breaking scale, while the local flow velocity near the barrier can be even larger. Pair-breaking excitations are therefore expected to become important, leading to a substantial suppression of the order-parameter amplitude and the subsequent breakdown of macroscopic phase coherence. Consequently, the phase of the order parameter becomes ill-defined, causing $\nu_{\mathrm{W}}$ to exhibit the observed drop and irregular behavior, signaling the collapse of the Josephson effect.

This observation of synchronized soliton emission prior to the pair-breaking breakdown is consistent with previous experimental observations in box traps~\cite{bernhartObservationShapiroSteps2025a,delpaceShapiroStepsStronglyinteracting2025a} and with a bosonic theoretical study~\cite{singhShapiroStepsDriven2024b}, where synchronized generation of solitons or vortex-antivortex pairs was reported.

From Figs.~\ref{fig:amp_phase_ufg}\textcolor{refcolor}{(c)} and \ref{fig:amp_phase_ufg}\textcolor{refcolor}{(f)}, we also find that the soliton-emission events occur approximately at $t = 2n \pi/\omega~(n=0,1,2,\ldots)$, when the barrier velocity $\dot{x}_{\mathrm{b}}$ reaches its maximum. This indicates that soliton generation is triggered when the magnitude of the local  flow velocity $|v_{\mathrm{b}}(t)|$ near the barrier reaches $v_{\mathrm{lc}}$, which occurs close to the time at which the barrier velocity is maximal.
Based on these observations, the fundamental mechanism of the Shapiro steps in a ring-trapped superfluid can be understood as a macroscopic phase-locking process governed by transitions of topological winding numbers: the ac modulation of the potential periodically drives $|v_{\mathrm{b}}(t)|$ at the barrier beyond the local critical velocity $v_{\mathrm{lc}}$, triggering discrete phase slips that incrementally drive the system between distinct metastable persistent current states characterized by integer winding numbers.

%%%%%%%%%%%%%%%%%%%%%%%%%%%%%%%%%%%%%%%%%%%%%%%%%%%%%%%%%%%%%%%%%%%%%%%%%%%%%%%%%%%%%%%%%%%%%%%%%%%%%%
%%%%%%%%%%%%%%%%%%%%%%%%%%%%%%%%%%%%%%%%%%%%%%%%%%%%%%%%%%%%%%%%%%%%%%%%%%%%%%%%%%%%%%%%%%%%%%%%%%%%%%
%%%%%%%%%%%%%%%%%%%%%%%%%%%%%%%%%%%%%%%%%%%%%%%%%%%%%%%%%%%%%%%%%%%%%%%%%%%%%%%%%%%%%%%%%%%%%%%%%%%%%%
\section{Conclusion} \label{sec:conclusion}
%%%%%%%%%%%%%%%%%%%%%%%%%%%%%%%%%%%%%%%%%%%%%%%%%%%%%%%%%%%%%%%%%%%%%%%%%%%%%%%%%%%%%%%%%%%%%%%%%%%%%%
%%%%%%%%%%%%%%%%%%%%%%%%%%%%%%%%%%%%%%%%%%%%%%%%%%%%%%%%%%%%%%%%%%%%%%%%%%%%%%%%%%%%%%%%%%%%%%%%%%%%%%
%%%%%%%%%%%%%%%%%%%%%%%%%%%%%%%%%%%%%%%%%%%%%%%%%%%%%%%%%%%%%%%%%%%%%%%%%%%%%%%%%%%%%%%%%%%%%%%%%%%%%%

  We have investigated the transport properties of a superfluid Fermi gas confined in a ring trap with a moving potential barrier across the BCS-BEC crossover. By employing the tdBdG equations, we simulated a JJ biased by combined dc and ac currents and  identified low-order Shapiro-step-like plateaus whose fitted levels lie close to integer multiples of $\hbar\omega/2$ across the  interaction range considered, provided that phase coherence is maintained. 
  A central feature of the ring geometry is its ability to support multiple metastable persistent current states, uniquely characterized by the topological winding number. 
  Our microscopic analysis reveals that the fundamental mechanism of the Shapiro steps can be understood as a process of synchronized phase-locking, where the system periodically increments its winding number in response to the potential modulation. 
  This synchronized dynamics is mediated by the periodic generation of solitons at the barrier. The representative emitted soliton analyzed in Fig.~\ref{fig:soliton_profiles} exhibits an approximately $\pi$ local phase jump, while its nucleation is associated with a global $2\pi$ phase-slip event that changes the winding number.
  These findings provide a microscopic perspective for understanding nonequilibrium transport in fermionic superfluids and offer important insights for the development of diverse atomtronic circuits in nontrivial topologies.
  Finally, we outline several promising directions for future research. While the present study focuses on an effective one-dimensional model to elucidate the fundamental phase-locking mechanisms, extending our numerical analysis to two- or three-dimensional ring geometries remains an important next step. Such higher-dimensional simulations 
  would make it possible to examine the snake instability, through which solitons periodically generated at the barrier may decay into vortex-antivortex pairs or vortex rings. Furthermore, exploring the dynamics beyond the mean-field BdG framework will be essential to fully capture the complex, incoherent transport processes in the strongly interacting BEC regime, where the validity of the simple phase-locking picture is fundamentally tested.

%%%%%%%%%%%%%%%%%%%%%%%%%%%%%%%%%%%%%%%%%%%%%%%%%%%%%%%%%%%%%%%%%%%%%%%%%%%%%%%%%%%%%%%%%%%%%%%%%%%%%%
%%%%%%%%%%%%%%%%%%%%%%%%%%%%%%%%%%%%%%%%%%%%%%%%%%%%%%%%%%%%%%%%%%%%%%%%%%%%%%%%%%%%%%%%%%%%%%%%%%%%%%
%%%%%%%%%%%%%%%%%%%%%%%%%%%%%%%%%%%%%%%%%%%%%%%%%%%%%%%%%%%%%%%%%%%%%%%%%%%%%%%%%%%%%%%%%%%%%%%%%%%%%%
\section*{Acknowledgments} \label{sec:acknowledgments}
%%%%%%%%%%%%%%%%%%%%%%%%%%%%%%%%%%%%%%%%%%%%%%%%%%%%%%%%%%%%%%%%%%%%%%%%%%%%%%%%%%%%%%%%%%%%%%%%%%%%%%
%%%%%%%%%%%%%%%%%%%%%%%%%%%%%%%%%%%%%%%%%%%%%%%%%%%%%%%%%%%%%%%%%%%%%%%%%%%%%%%%%%%%%%%%%%%%%%%%%%%%%%
%%%%%%%%%%%%%%%%%%%%%%%%%%%%%%%%%%%%%%%%%%%%%%%%%%%%%%%%%%%%%%%%%%%%%%%%%%%%%%%%%%%%%%%%%%%%%%%%%%%%%%

This work was supported by JSPS KAKENHI Grant No.~JP25K00215 (M.K.) and JST ASPIRE No.~JPMJAP24C2 (M.K.).

%%%%%%%%%%%%%%%%%%%%%%%%%%%%%%%%%%%%%%%%%%%%%%%%%%%%%%%%%%%%%%%%%%%%%%%%%%%%%%%%%%%%%%%%%%%%%%%%%%%%%%
%%%%%%%%%%%%%%%%%%%%%%%%%%%%%%%%%%%%%%%%%%%%%%%%%%%%%%%%%%%%%%%%%%%%%%%%%%%%%%%%%%%%%%%%%%%%%%%%%%%%%%
%%%%%%%%%%%%%%%%%%%%%%%%%%%%%%%%%%%%%%%%%%%%%%%%%%%%%%%%%%%%%%%%%%%%%%%%%%%%%%%%%%%%%%%%%%%%%%%%%%%%%%
\section*{Data Availability} \label{sec:data_availability}
%%%%%%%%%%%%%%%%%%%%%%%%%%%%%%%%%%%%%%%%%%%%%%%%%%%%%%%%%%%%%%%%%%%%%%%%%%%%%%%%%%%%%%%%%%%%%%%%%%%%%%
%%%%%%%%%%%%%%%%%%%%%%%%%%%%%%%%%%%%%%%%%%%%%%%%%%%%%%%%%%%%%%%%%%%%%%%%%%%%%%%%%%%%%%%%%%%%%%%%%%%%%%
%%%%%%%%%%%%%%%%%%%%%%%%%%%%%%%%%%%%%%%%%%%%%%%%%%%%%%%%%%%%%%%%%%%%%%%%%%%%%%%%%%%%%%%%%%%%%%%%%%%%%%
The data that support the findings of this article are publicly available~\cite{kuriki_2026_22240700}.

%%%%%%%%%%%%%%%%%%%%%%%%%%%%%%%%%%%%%%%%%%%%%%%%%%%%%%%%%%%%%%%%%%%%%%%%%%%%%%%%%%%%%%%%%%%%%%%%%%%%%%
%%%%%%%%%%%%%%%%%%%%%%%%%%%%%%%%%%%%%%%%%%%%%%%%%%%%%%%%%%%%%%%%%%%%%%%%%%%%%%%%%%%%%%%%%%%%%%%%%%%%%%
%%%%%%%%%%%%%%%%%%%%%%%%%%%%%%%%%%%%%%%%%%%%%%%%%%%%%%%%%%%%%%%%%%%%%%%%%%%%%%%%%%%%%%%%%%%%%%%%%%%%%%
\appendix
%%%%%%%%%%%%%%%%%%%%%%%%%%%%%%%%%%%%%%%%%%%%%%%%%%%%%%%%%%%%%%%%%%%%%%%%%%%%%%%%%%%%%%%%%%%%%%%%%%%%%%
%%%%%%%%%%%%%%%%%%%%%%%%%%%%%%%%%%%%%%%%%%%%%%%%%%%%%%%%%%%%%%%%%%%%%%%%%%%%%%%%%%%%%%%%%%%%%%%%%%%%%%
%%%%%%%%%%%%%%%%%%%%%%%%%%%%%%%%%%%%%%%%%%%%%%%%%%%%%%%%%%%%%%%%%%%%%%%%%%%%%%%%%%%%%%%%%%%%%%%%%%%%%%

%%%%%%%%%%%%%%%%%%%%%%%%%%%%%%%%%%%%%%%%%%%%%%%%%%%%%%%%%%%%%%%%%%%%%%%%%%%%%%%%%%%%%%%%%%%%%%%%%%%%%%
%%%%%%%%%%%%%%%%%%%%%%%%%%%%%%%%%%%%%%%%%%%%%%%%%%%%%%%%%%%%%%%%%%%%%%%%%%%%%%%%%%%%%%%%%%%%%%%%%%%%%%
%%%%%%%%%%%%%%%%%%%%%%%%%%%%%%%%%%%%%%%%%%%%%%%%%%%%%%%%%%%%%%%%%%%%%%%%%%%%%%%%%%%%%%%%%%%%%%%%%%%%%%
\section{Modified Broyden's method} \label{sec:modified_broyden}
%%%%%%%%%%%%%%%%%%%%%%%%%%%%%%%%%%%%%%%%%%%%%%%%%%%%%%%%%%%%%%%%%%%%%%%%%%%%%%%%%%%%%%%%%%%%%%%%%%%%%%
%%%%%%%%%%%%%%%%%%%%%%%%%%%%%%%%%%%%%%%%%%%%%%%%%%%%%%%%%%%%%%%%%%%%%%%%%%%%%%%%%%%%%%%%%%%%%%%%%%%%%%
%%%%%%%%%%%%%%%%%%%%%%%%%%%%%%%%%%%%%%%%%%%%%%%%%%%%%%%%%%%%%%%%%%%%%%%%%%%%%%%%%%%%%%%%%%%%%%%%%%%%%%

\begin{figure}[htbp]
  \centering
  \includegraphics[width=1.0\linewidth]{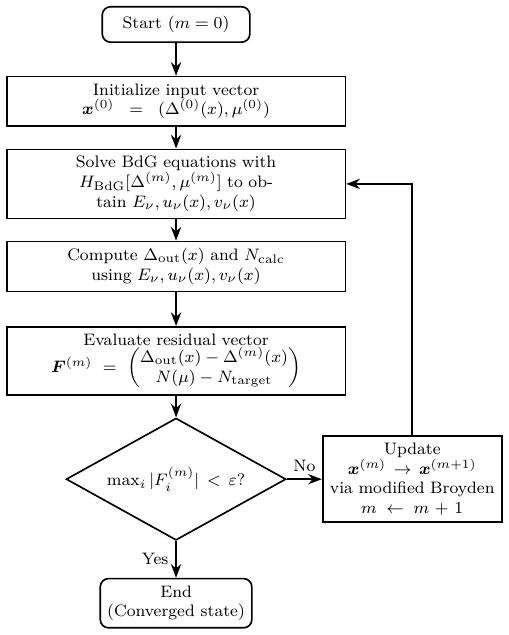}
  \caption{Flowchart of the self-consistent calculation loop using the modified Broyden's method to solve the BdG equations. We set the convergence criterion to $\varepsilon = 10^{-8}$ for the maximum component of the residual vector $\bm{F}^{(m)}$.}
  \label{fig:broyden_flowchart}
\end{figure}

In this appendix, we outline the modified Broyden's method~\cite{johnsonModifiedBroydensMethod1988} used to solve the BdG equations, the gap equation, and the particle number equation [Eqs.~\eqref{eq:BdG_1D}--\eqref{eq:particle_number_1D}] self-consistently. The overall procedure of this self-consistent calculation loop is summarized in the flowchart shown in Fig.~\ref{fig:broyden_flowchart}.
This efficient quasi-Newton approach is widely adopted across various physical disciplines, ranging from condensed matter~\cite{giannozziQUANTUMESPRESSOModular2009} to nuclear structure calculations~\cite{baranBroydensMethodNuclear2008}. In this method, one can update the inverse Jacobian of the residual function without the computational overhead of storing or inverting large $D \times D$ matrices (where $D$ is the number of degrees of freedom).

Let $\bm{x}^{(m)}$ be the generalized real vector to be optimized at the $m$th iteration, and let $\bm{F}(\bm{x}^{(m)})$ be the residual vector corresponding to $\bm{x}^{(m)}$. The goal is to find the root $\bm{F}(\bm{x}) = \bm{0}$.  
The update rule for the $(m+1)$st iteration is given by
\begin{align}
  \bm{x}^{(m+1)} = \bm{x}^{(m)} + \alpha \bm{F}^{(m)} - \sum_{n=\tilde{m}}^{m-1} w_n \gamma_{mn} \bm{u}^{(n)},
  \label{eq:broyden_update}
\end{align}
where $\alpha$ is a linear mixing parameter chosen to ensure stability and $\tilde{m} = \max(1, m-M)$, with $M$ being the number of stored history vectors. The vector $\bm{u}^{(n)}$ is defined as 
\begin{align}
  \bm{u}^{(n)} = \alpha \Delta \bm{F}^{(n)} + \Delta \bm{x}^{(n)},
\end{align}
where we employ the normalized differences given by
\begin{align}
  \Delta \bm{F}^{(n)} = \frac{\bm{F}^{(n+1)} - \bm{F}^{(n)}}{|\bm{F}^{(n+1)} - \bm{F}^{(n)}|}, \label{eq:delta_F} 
\end{align}
and
\begin{align}
  \Delta \bm{x}^{(n)} = \frac{\bm{x}^{(n+1)} - \bm{x}^{(n)}}{|\bm{F}^{(n+1)} - \bm{F}^{(n)}|}. \label{eq:delta_x} 
\end{align}
The coefficients $\gamma_{mn}$ account for the iteration history and are computed by
\begin{align}
  \gamma_{mn} = \sum_{k=\tilde{m}}^{m-1} c_k^{(m)} \beta_{kn},
\end{align}
where $c_k^{(m)} = w_k \langle \Delta \bm{F}^{(k)} | \bm{F}^{(m)} \rangle$. Here, $\langle \cdot | \cdot \rangle$ denotes the inner product. The matrix $\boldsymbol{\beta}$ is the inverse of the small $M \times M$ matrix defined as
\begin{align}
  \boldsymbol{\beta} = (w_0^2 \bm{I} + \bm{a})^{-1},
\end{align}
where the elements of the matrix $\bm{a}$ are given by the inner products of the residual differences
\begin{align}
  a_{ij} = w_i w_j \langle \Delta \bm{F}^{(j)} | \Delta \bm{F}^{(i)} \rangle.
\end{align}
In these expressions, $w_n$ are weights assigned to each previous iteration to improve stability. 

This algorithm allows for the accumulation of information from previous iterations to construct a better approximation of the inverse Jacobian. In practice, this modified scheme effectively reduces the number of iterations required for convergence compared with the ordinary Broyden's method~\cite{broydenClassMethodsSolving1965a}, offering a distinct advantage in both speed and memory usage.

In our implementation, we construct $\bm{x}$ as an $(N_x+1)$-dimensional vector consisting of the order parameter $\Delta(x_i)$ at the $N_x$ spatial grid points and the chemical potential $\mu$. In the stationary initial state considered here, the order parameter can be chosen real by a global gauge transformation. Accordingly, the residual vector $\bm{F}(\bm{x})$ is constructed such that its spatial components represent the difference between the output and input order parameters, $\Delta_{\mathrm{out}}(x_i) - \Delta_{\mathrm{in}}(x_i)$, while the final component represents the deviation of the calculated particle number from the target value, $N(\mu) - N_{\mathrm{target}}$. We set the number of stored history vectors to $M = 7$, the linear mixing parameter to $\alpha = 0.5$, the inverse Jacobian update constraint weight to $w_0 = 0.01$, and the iteration weights to $w_n = 1$ for $n \ge 1$. For the calculation of the initial stationary states, we regard convergence as achieved when all components of the residual vector $\bm{F}(\bm{x})$ are smaller than $ \varepsilon = 10^{-8}$ in magnitude.

%%%%%%%%%%%%%%%%%%%%%%%%%%%%%%%%%%%%%%%%%%%%%%%%%%%%%%%%%%%%%%%%%%%%%%%%%%%%%%%%%%%%%%%%%%%%%%%%%%%%%%
%%%%%%%%%%%%%%%%%%%%%%%%%%%%%%%%%%%%%%%%%%%%%%%%%%%%%%%%%%%%%%%%%%%%%%%%%%%%%%%%%%%%%%%%%%%%%%%%%%%%%%
\section{Calculation of the effective charging energy} \label{sec:charging_energy}
%%%%%%%%%%%%%%%%%%%%%%%%%%%%%%%%%%%%%%%%%%%%%%%%%%%%%%%%%%%%%%%%%%%%%%%%%%%%%%%%%%%%%%%%%%%%%%%%%%%%%%
%%%%%%%%%%%%%%%%%%%%%%%%%%%%%%%%%%%%%%%%%%%%%%%%%%%%%%%%%%%%%%%%%%%%%%%%%%%%%%%%%%%%%%%%%%%%%%%%%%%%%%

In this appendix, we explain the calculation of the effective charging energy $E_{\mathrm{C}}$ in Sec.~\ref{sec:potential_barrier}. 
The effective charging energy of the Josephson junction is defined as
\begin{align}
  E_{\mathrm{C}} = 4 \left( \frac{\partial \mu}{\partial N} \right)_{\mathcal{V}, a}. \label{eq:charging_energy_def}
\end{align}
To properly evaluate this thermodynamic derivative, it is crucial to keep the physical volume $\mathcal{V}$ and the $s$-wave scattering length $a$ strictly constant.
To compute the chemical potential $\mu$ as a function of the total particle number $N$ across various interaction regimes, we solve the time-independent BdG equations [Eqs.~\eqref{eq:BdG_1D}--\eqref{eq:particle_number_1D}] self-consistently for a uniform system ($V_0 = 0$) for each $N$ around the target value $N_0 = 2000$.
Throughout this section, quantities with the subscript $0$ (e.g., $N_0$, $E_{\mathrm{F}0}$, $L_{i0}$) denote the corresponding quantities evaluated at the target particle number $N_0$.
In these calculations, the convergence criterion for the residual vector is relaxed to $10^{-4}$ in magnitude, unlike the stricter condition described in Appendix~\ref{sec:modified_broyden}.

In our numerical scheme, we strictly enforce the dimensionless density to be fixed at $\tilde{n} = 1/(3\pi^2)$ for all $N$. This requires the dimensionless system sizes to be dynamically scaled as $L_i(N) = L_{i0} (N/N_0)^{1/3}$ ($i = x, y, z$). Accordingly, to maintain a constant physical scattering length $a$ and cutoff energy $E_{\mathrm{cut}}$, the interaction parameter and the cutoff energy used as input to the BdG equations must also be scaled as $[k_{\mathrm{F}}(N)a]^{-1} = [k_{\mathrm{F0}}a]^{-1} (N_0/N)^{1/3}$ and $E_{\mathrm{cut}}(N) = E_{\mathrm{cut}0} (N_0/N)^{2/3}$, respectively. Here, $L_i(N)$ and $E_{\mathrm{cut}}(N)$ denote the numerical input values in units of $k_{\mathrm{F}}(N)^{-1}$ and $E_{\mathrm{F}}(N)$, respectively; the corresponding physical quantities are held fixed.

\begin{figure}[htbp]
  \centering
  \includegraphics[width=1.0\linewidth]{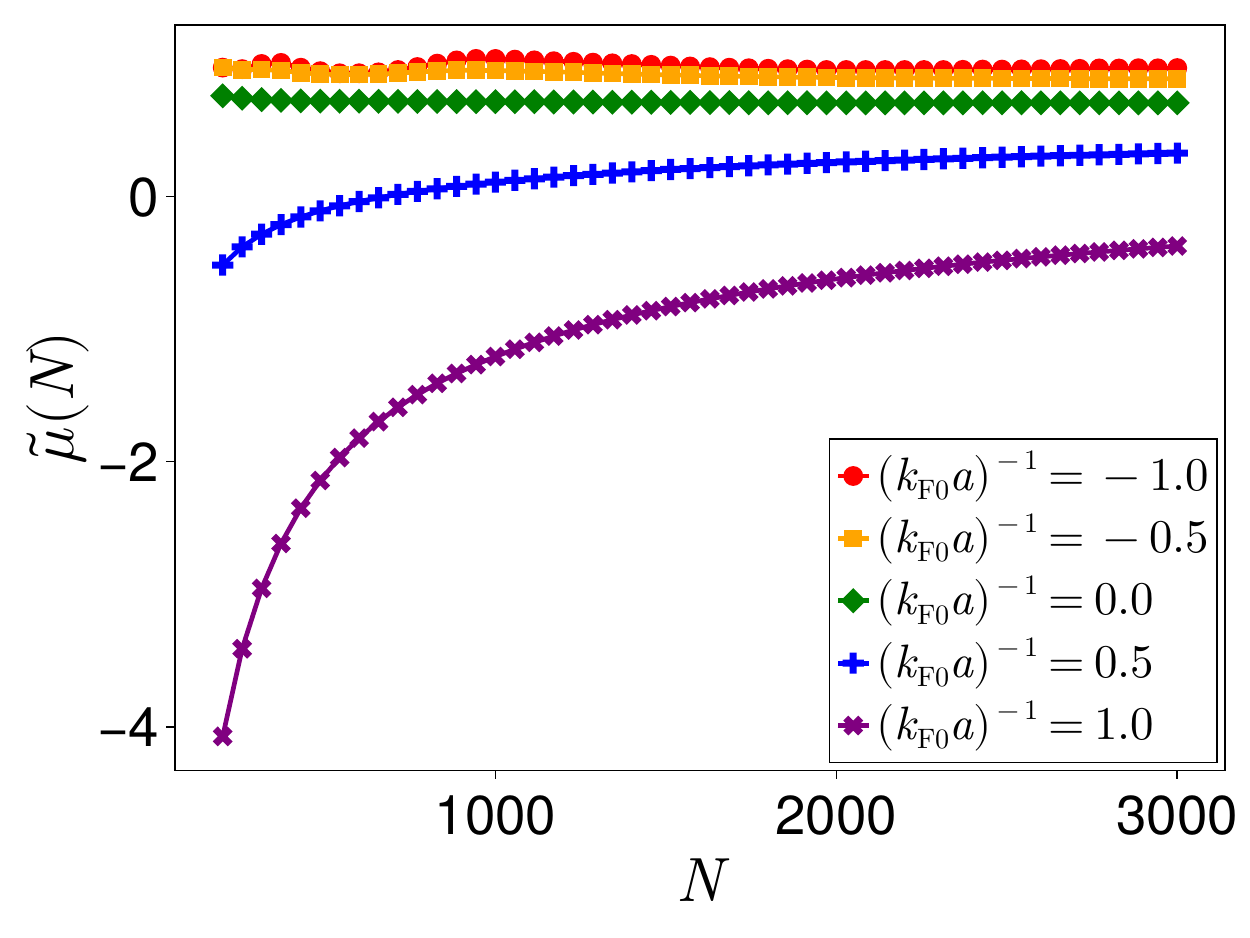}
  \caption{Scaled chemical potential $\tilde{\mu}(N) = \mu(N)/E_{\mathrm{F}}(N)$ as a function of the total particle number $N$ for various interaction strengths $[k_{\mathrm{F}0}a]^{-1}$ at the target particle number $N_0 = 2000$. The data are obtained by dynamically scaling the system parameters to maintain a constant physical volume and scattering length.}
  \label{fig:chempot_N_ka_inv}
\end{figure}

Figure~\ref{fig:chempot_N_ka_inv} shows the resulting $N$ dependence of the scaled chemical potential $\tilde{\mu}(N) \equiv \mu(N)/E_{\mathrm{F}}(N)$ for different interaction regimes. 
To accurately evaluate the derivative from the discrete numerical data, we apply a Savitzky--Golay filter~\cite{savitzkySmoothingDifferentiationData1964} with a window size of 11 points and a polynomial order of 3, followed by a cubic spline interpolation. 

Since the Fermi energy itself depends on the particle number as $E_{\mathrm{F}}(N) = E_{\mathrm{F0}}(N/N_0)^{2/3}$, the physical derivative must account for this scaling. The correct derivative at the target particle number $N_0$ is analytically derived as
\begin{align}
  \frac{1}{E_{\mathrm{F}0}} \left. \frac{\partial \mu}{\partial N} \right|_{N=N_0} &= \left. \frac{\partial \tilde{\mu}(N)}{\partial N} \right|_{N=N_0} + \frac{2}{3N_0} \tilde{\mu}(N_0), \label{eq:derivative_conversion}
\end{align}
where the second term precisely corrects for the apparent variation caused by the continuous rescaling of the energy unit $E_{\mathrm{F}}(N)$.
Finally, by substituting this corrected derivative into Eq.~\eqref{eq:charging_energy_def}, we determine the precise value of $E_{\mathrm{C}}$ for each interaction strength.

%%%%%%%%%%%%%%%%%%%%%%%%%%%%%%%%%%%%%%%%%%%%%%%%%%%%%%%%%%%%%%%%%%%%%%%%%%%%%%%%%%%%%%%%%%%%%%%%%%%%%%
%%%%%%%%%%%%%%%%%%%%%%%%%%%%%%%%%%%%%%%%%%%%%%%%%%%%%%%%%%%%%%%%%%%%%%%%%%%%%%%%%%%%%%%%%%%%%%%%%%%%%%
%%%%%%%%%%%%%%%%%%%%%%%%%%%%%%%%%%%%%%%%%%%%%%%%%%%%%%%%%%%%%%%%%%%%%%%%%%%%%%%%%%%%%%%%%%%%%%%%%%%%%%
\section{Split-step Fourier method} \label{sec:split_step_fourier}
%%%%%%%%%%%%%%%%%%%%%%%%%%%%%%%%%%%%%%%%%%%%%%%%%%%%%%%%%%%%%%%%%%%%%%%%%%%%%%%%%%%%%%%%%%%%%%%%%%%%%%
%%%%%%%%%%%%%%%%%%%%%%%%%%%%%%%%%%%%%%%%%%%%%%%%%%%%%%%%%%%%%%%%%%%%%%%%%%%%%%%%%%%%%%%%%%%%%%%%%%%%%%
%%%%%%%%%%%%%%%%%%%%%%%%%%%%%%%%%%%%%%%%%%%%%%%%%%%%%%%%%%%%%%%%%%%%%%%%%%%%%%%%%%%%%%%%%%%%%%%%%%%%%%

In this appendix, we explain the split-step Fourier method~\cite{reichlCoreFillingSnaking2017a,tokimotoExcitationHiggsMode2019b,wangCollisionsMajoranaZero2023a} used to solve the tdBdG equations and the time-dependent gap equation [Eqs.~\eqref{eq:tdBdG_1D} and \eqref{eq:tdgap_eq_1D}].

First, we can rewrite Eq.~\eqref{eq:tdBdG_1D} as
\begin{align}
  i\hbar \frac{\partial}{\partial t} \Psi_{\nu}(x,t) 
  &= H_{\mathrm{BdG}}(t) \Psi_{\nu}(x,t) \notag\\
  &= [H_{\mathrm{K}} + H_{\mathrm{P}}(t)] \Psi_{\nu}(x,t), \label{eq:tdBdG_split}
\end{align}
where we define $\Psi_{\nu}(x,t) = (u_{\nu}(x,t), v_{\nu}(x,t))^{\mathrm{T}}$ and 
\begin{align}
  H_{\mathrm{BdG}}(t) &= H_{\mathrm{K}} + H_{\mathrm{P}}(t), \\
  H_{\mathrm{K}} &= 
  \begin{pmatrix}
    -\frac{\hbar^2}{2m} \frac{\partial^2}{\partial x^2} + i\hbar v_{\mathrm{DC}} \frac{\partial}{\partial x} & 0 \\
    0 & \frac{\hbar^2}{2m} \frac{\partial^2}{\partial x^2} + i\hbar v_{\mathrm{DC}} \frac{\partial}{\partial x}
  \end{pmatrix}, \\
  H_{\mathrm{P}}(t) &=
  \begin{pmatrix}
    E_{\perp} + V(x,t) - \mu & \Delta(x,t) \\
    \Delta^*(x,t) & -E_{\perp} - V(x,t) + \mu
  \end{pmatrix}.
\end{align}
Here, $E_{\perp} = \hbar^2 (k_y^2 + k_z^2)/(2m)$ is the transverse kinetic energy.
We then formally write the time evolution from $t$ to $t+\Delta t$ as
\begin{align}
  &\Psi_{\nu}(x,t+\Delta t)\notag\\
  &= \mathcal{T}\exp\left\{ \left[ -\frac{i}{\hbar}\int_t^{t+\Delta t} dt'~ H_{\mathrm{BdG}}(t')\right]\right\} \Psi_{\nu}(x,t), \label{eq:time_evolution}
\end{align}
where $\mathcal{T}$ is the time-ordering operator.
Applying the time-ordered exponential splitting method to Eq.~\eqref{eq:time_evolution}, we obtain
\begin{align}
  \Psi_{\nu}(x,t+\Delta t) \simeq & \exp\left[-\frac{i}{2\hbar} H_{\mathrm{P}}(t+\Delta t/2) \Delta t\right] \notag\\
  &\times \exp\left[-\frac{i}{\hbar} H_{\mathrm{K}} \Delta t\right] \notag\\
  &\times \exp\left[-\frac{i}{2\hbar} H_{\mathrm{P}}(t+\Delta t/2) \Delta t\right] \Psi_{\nu}(x,t). \label{eq:time_evolution_split}
\end{align}
In the actual implementation, the midpoint order parameter $\Delta (x, t+\Delta t/2)$ appearing in $H_{\mathrm{P}}(t+\Delta t/2)$ is approximated by the self-consistent value at the beginning of the time step, $\Delta (x,t)$. Since the order parameter evolves smoothly over a single time step for the parameters considered here, this approximation remains sufficiently accurate in the present simulations.
Since all matrix elements of $H_{\mathrm{P}}(t)$ are local in real space, the exponential of this term can be computed directly as
\begin{align}
  & e^{-i H_{\mathrm{P}}(t) \Delta t/\hbar}\notag\\
  & = \frac{1}{[\epsilon(x,t) - \eta(x,t)]^2 + |\Delta(x,t)|^2}
  \begin{pmatrix}
    \alpha(x,t) & \beta(x,t) \\
    -\beta^*(x,t) & \alpha^*(x,t)
  \end{pmatrix}
  , \label{eq:exp_H_P}
\end{align}
where $ \eta(x,t) = E_{\perp} + V(x,t) - \mu$, $\epsilon (x,t) = [\eta (x,t)^2 + |\Delta (x,t)|^2]^{1/2} $, and
\begin{align}
  \alpha(x,t) =& |\Delta(x,t)|^2 e^{-i \epsilon(x,t) \Delta t/\hbar} \notag\\ 
         & + [\epsilon(x,t) - \eta(x,t)]^2 e^{i\epsilon(x,t) \Delta t/\hbar}, \\
  \beta(x,t) =& -2i \Delta(x,t) [\epsilon(x,t) - \eta(x,t)] \sin [\epsilon(x,t) \Delta t/\hbar].
\end{align}
$H_{\mathrm{K}}$ has a diagonal representation in momentum space. Therefore, we can efficiently compute the time evolution by switching between real and momentum spaces using the fast Fourier transform algorithm. We calculate the time evolution of all quasiparticle wave functions $\Psi_{\nu}(x,t)$ satisfying $E_\nu \ge 0$ for every $(k_y, k_z)$ mode, and we recalculate the order parameter $\Delta(x,t)$ self-consistently at each time step using Eq.~\eqref{eq:tdgap_eq_1D}.

% REVTeX selects apsrev4-2 and writes the longbibliography controls.
\bibliography{Shapirosteps}

%\tableofcontents
\end{document}